\documentclass[aps,pra,reprint,amsfonts,amssymb,superscriptaddress,floatfix,eprint,longbibliography]{revtex4-1}
\usepackage{graphicx}
\usepackage{amsmath}
\usepackage{dcolumn}
\usepackage{color}
\usepackage{calc}
\usepackage{MnSymbol}
\usepackage{hyperref}
\usepackage{hypernat}
\usepackage{enumitem}
\usepackage{float}
\usepackage{relsize}
\hypersetup{colorlinks=true, citecolor=magenta, urlcolor=blue}

\begin{document}

\title{Stationary states, dynamical stability, and vorticity of Bose-Einstein condensates in tilted rotating harmonic traps}
\author{Srivatsa B. Prasad}
\email{srivatsa.badariprasad@unimelb.edu.au}
\affiliation{School of Physics, University of Melbourne, Melbourne, 3010, Australia}
\author{Brendan C. Mulkerin}
\affiliation{Centre for Quantum and Optical Science, Swinburne University of Technology, Melbourne, 3122, Australia}
\author{Andrew M. Martin}
\affiliation{School of Physics, University of Melbourne, Melbourne, 3010, Australia}

\date{\today}

\begin{abstract}
We theoretically investigate a Bose-Einstein condensate confined by a rotating harmonic trap whose rotation axis is not aligned with any of its principal axes. The principal axes of the Thomas-Fermi density profiles of the resulting stationary solutions are found to be tilted with respect to those of the rotating trap, representing an extra degree of freedom that is associated with the existence of additional branches of stationary solutions for any given rotation axis alignment. By linearizing the time-dependent theory about the stationary states, we obtain a semi-analytical prediction of their dynamical instability at high rotation frequencies against collective modes arising from environmental perturbations. Comparing the stationary states to direct simulations of the Gross-Pitaevskii equation, we predict the nucleation of quantum vortices in the dynamically unstable rotational regime. These vortex lines are aligned along the rotation axis despite the tilting of the rotating trap although the background density profile is tilted with respect to the trapping and rotation axes.
\end{abstract}

\maketitle
\section{\label{sec:level0}Introduction}
For more than 70 years the behaviour of rotating superfluids have been of considerable experimental and theoretical interest to several generations of physicists. Initially, it was recognized by Onsager~\cite{nuovocimento_6_supp2_279-287_1949} and Feynman~\cite{proglowtempphys_1_2_17-53_1955} that superfluid flow is characterized by a quantized circulation, which was subsequently experimentally verified by Hall and Vinen \cite{procrsoca_238_1213_204-214_1956,procrsoca_238_1213_215-234_1956}. Since then, it has been recognized that the presence of a nonzero angular momentum in a superfluid leads to complex, nontrivial behaviour, spurring discoveries such as the recent realization of negative-temperature Onsager vortex clusters in two-dimensional Bose gases~\cite{science_364_6447_1264-1267_2019, science_364_6447_1267-1271_2019}. In particular, Bose-Einstein condensates (BECs) offer a uniquely flexible platform for the study of superfluid rotation and have been the focus of intense study for several years~\cite{pitaevskiistringaribec, advphys_57_6_539-616_2008, rmp_81_2_647-691_2009}.

Consider a superfluid in a rotating bucket. This superfluid cannot support rigid-body rotation if the bucket is symmetric about the rotation axis, due to the absence of shear forces, and so its angular momentum manifests in the form of quantum vortices above a certain critical rotation frequency~\cite{leggettquantumliquids}. However, when the symmetry of the bucket about the rotation axis is broken, it is able to transfer angular momentum to the superfluid even in the absence of shear forces, and the condensate exhibits solid-body rotation at slow rotation frequencies and quantized vortices above a critical rotation frequency. For Bose-Einstein condensates, in which the `bucket' is replaced by an atomic trap generated by electromagnetic fields, angular momentum may be transferred from the trap to the condensate by modulating the applied fields such that the condensate is confined by a rotating potential that is asymmetrical about the rotation axis~\cite{pitaevskiistringaribec}. At sufficiently high rotation frequencies, this method induces vorticity in the condensate~\cite{prl_88_1_010405_2001}. Alternate methods for producing vortices in BECs also exist, such as stirring with a Gaussian laser beam~\cite{prl_84_5_806-809_2000}, dragging a laser configuration through a trapped condensate (or, equivalently, a condensate through a laser configuration)~\cite{prl_104_16_160401_2010, prl_117_24_245301_2016}, applying oscillatory perturbations to the trapping~\cite{pra_79_4_043618_2009,prl_103_4_045301_2009}, condensing a rotating thermal (non-condensed) atomic vapor~\cite{prl_87_23_210403_2001}, and utilizing the Kibble-Zurek mechanism by quenching a thermal vapor across the BEC critical temperature~\cite{nature_455_7215_948-951_2008}. The theoretical and analytical study of the resulting vortices have uncovered phenomena rich in variety; some examples that are relevant to scalar, nondipolar, single-component condensates at zero temperature includes Kelvin waves~\cite{pra_62_6_063617_2000, prl_90_10_100403_2003, prl_101_2_020402_2008}, Abrikosov vortex lattices and their Tkachenko modes~\cite{science_292_5516_476-479_2001, prl_91_11_110402_2003, prl_91_10_100402_2003}, quantum Hall-like physics~\cite{prl_87_6_060403_2001, prl_87_12_120405_2001, prl_91_3_030402_2003, prl_92_4_040404_2004}, vortex reconnections~\cite{physfluids_24_1_125108_2012, prx_7_2_021031_2017}, quantum analogs of classical fluid instabilities~\cite{prl_104_15_150404_2010, prl_117_24_245301_2016, pra_97_5_053608_2018}, and hysteresis~\cite{pra_63_4_041603r_2001, pra_74_4_043618_2006}.

By contrast, previous studies relating to tilting effects in rotating BECs have mainly focussed on the collective modes of vortices in response to tilting perturbations of the trap~\cite{prl_86_21_4725-7428_2001, prl_91_9_090403_2003, prl_93_8_080406_2004, prl_113_16_165303_2014}, while the literature concerning the steady rotation of the external confinement about a non-principal axis is chiefly limited to the stability of the centre-of-mass oscillations in the rotating frame~\cite{pra_65_6_063606_2002, pra_71_4_043610_2005}. Given that such \emph{tilted} rotating traps may be experimentally generated in a similar manner to the excitation of the tilting modes, and that roughly analogous systems such as dipolar BECs with tilted rotating dipole moments have been realized experimentally~\cite{prl_89_13_130401_2002, prl_120_23_230401_2018}, a systematic study of BECs confined by a tilted rotating trap is warranted. In this paper, we analytically obtain stationary solutions of the Gross-Pitaevskii equation in the Thomas-Fermi limit for a condensate subject to a range of different tilting angles and harmonic trapping regimes. For all but the most trivial cases, the stationary solution densities are found to be tilted about the rotation axis by a different angle than the trap itself. One of the consequences of this additional degree of freedom is the existence of two previously unknown branches of stationary solutions. These exist even when the trap is not tilted away from the rotation axis, a result that is analogous to the tilted triaxial ellipsoids that are rotating-frame stationary solutions for self-gravitating irrotational classical fluids~\cite{physfluids_8_12_3414-3422_1996}. Focusing on the stationary solution branch existing in the nonrotating limit, we semi-analytically linearize the condensate's fluctuations in response to small perturbations and thus predict a dynamical instability at higher rotation frequencies, where the amplitude of one or more collective modes is expected to amplify exponentially in time. In the regions of dynamical instability we show via numerical Gross-Pitaevskii simulations that untilted vortices are nucleated from a tilted condensate, despite the background condensate density still being tilted. The theoretical formalism utilized represents a generalization of existing theoretical methods for studying BECs in asymmetric, untilted rotating traps~\cite{prl_86_3_377-380_2001, prl_87_19_190402_2001, prl_92_2_020403_2004, prl_95_14_145301_2005, pra_73_6_061603r_2006, jphysb_40_18_3615-3628_2007} and are readily conducive to experimental investigation along the lines of previous studies that have probed the untilted regime~\cite{prl_86_20_4443-4446_2001, prl_88_1_010405_2001, prl_88_7_070406_2002}.

This paper is structured as follows. Section \ref{sec:level1} defines the concept of a tilted rotating trap and introduces the relevant coordinate reference frames, while Sec.~\ref{sec:level2} discusses the methodology for solving for the vortex-free stationary solutions in the Thomas-Fermi limit. In Sec.~\ref{sec:level3}, we examine the features of these stationary solutions for two distinct trapping regimes, and in Sec.~\ref{sec:level4}, the time-dependent theory is linearized in order to characterize the dynamical stability of the vortex-free stationary solutions during a quasi-adiabatic rampup of the trapping rotation frequency. Finally, Sec.~\ref{sec:level5} contains a discussion of the outcomes of a series of numerical simulations of such rampups, where the dynamical route to vortex nucleation in a vorticity-free condensate in a tilted rotating trap is demonstrated.

\section{\label{sec:level1}The Tilted, Rotating Harmonic Trap}
In order to describe a dilute, scalar BEC, at zero temperature in a rotating, tilted harmonic trap, we utilize the Gross-Pitaevskii equation for the condensate order parameter, $\psi$. We assume that $N$ condensed bosons, each with a mass $m$, are confined in the trap and that the root mean squared harmonic trapping frequency in the $x$-$y$ plane is given by $\omega_{\perp}$. This may be used to rescale $t$ as $t\rightarrow\omega_{\perp}t$ and $\mathbf{r}$ as $\mathbf{r}\rightarrow\mathbf{r}/l_{\perp}$, where $l_{\perp} = \sqrt{\hbar/(m\omega_{\perp})}$ is the in-plane harmonic oscillator length. We also rescale $\psi$ as $\psi\rightarrow\sqrt{l_{\perp}^3/N}\psi$, such that it is normalized as
\begin{equation}
  \int\mathrm{d}^3r\,|\Psi(\mathbf{r}, t)|^2 = 1. \label{eq:normalization}
\end{equation}
Subsequently, in a reference frame rotating with respect to the inertial laboratory frame with the angular velocity $\mathbf{\Omega}$, $\psi$ obeys the dimensionless Gross-Pitaevskii equation (GPE)~\cite{pitaevskiistringaribec, pethicksmithbecdilutegases, rmp_81_2_647-691_2009, advphys_57_6_539-616_2008}:
\begin{equation}
  i\frac{\partial\psi}{\partial t} = -\frac{1}{2}\nabla^2\psi + V_{\text{T}}(\mathbf{r},t)\psi + \tilde{g}|\psi|^2\psi + i\mathbf{\Omega}\cdot(\mathbf{r}\times\nabla)\psi. \label{eq:rescaledgpe}
\end{equation}
Here we define $\tilde{g} = 4\pi Na_{\text{s}}/l_{\perp}$ as the effective strength of a mean-field, two-body interaction, with a corresponding $s$-wave scattering length given by $a_s$, and denote the time-dependent harmonic trapping potential by $V_{\text{T}}$.

Previously, theoretical and experimental studies of the angular momentum of trapped BECs have tended to assume that the rotation axis of the confinement coincides with one of its symmetry axes. The rotation of a harmonic trap about an arbitrary axis can effectively be modeled by fixing $\mathbf{\Omega} = \Omega\hat{z}$, without loss of generality, and assuming that the trapping potential is not symmetric under the transformation $z \rightarrow -z$. In the co-rotating reference frame, this potential can be specified by
\begin{align}
V_{\text{T}}\left(\mathbf{r}\right) &= \frac{1}{2}\left[(1-\varepsilon)\left(x\cos\theta + z\sin\theta\right)^2 + (1+\varepsilon)y^2\right] \nonumber \\
&+ \frac{1}{2}\gamma^2\left(x\sin\theta - z\cos\theta\right)^2, \label{eq:tiltedtrapuprightcoord}
\end{align}
where $\varepsilon \in (-1, 1)$ and $\gamma\in\mathbb{R}$. This external potential is equivalent to
\begin{equation}
V_{\text{T}}(\mathbf{R}) = \frac{1}{2}\left[(1-\varepsilon)X^2 + (1+\varepsilon)Y^2 + \gamma^2Z^2\right], \label{eq:tiltedtrapnocrossterm}
\end{equation}
via a rotation of the co-rotating coordinates as given by
\begin{equation}
  \begin{pmatrix}
  X \\
  Y \\
  Z
  \end{pmatrix}
  =
\begin{pmatrix}
\cos\theta & 0 & \sin\theta \\
0 & 1 & 0 \\
-\sin\theta & 0 & \cos\theta
\end{pmatrix}
\begin{pmatrix}
x \\
y \\
z
\end{pmatrix}. \label{eq:traptilttransform}
\end{equation}
By inspection, Eq.~\eqref{eq:tiltedtrapuprightcoord} is equivalent to Eq.~\eqref{eq:tiltedtrapnocrossterm} when, for integer $n$, the tilting angle obeys $\theta = n\pi$. A similar equivalence, albeit with modified values of $\gamma$ and $\varepsilon$, holds when $n$ takes on half-integer values.

To simulate the stationary state and dynamics of a BEC in this trap via numerical methods, it is sufficient to use Eqs.~\eqref{eq:rescaledgpe} and \eqref{eq:tiltedtrapuprightcoord}. However, the vorticity-free stationary solutions of Eq.~\eqref{eq:tiltedtrapuprightcoord}, and their linear response to environmental perturbations, are well-described in the $\theta = 0$ limit by purely semi-analytical methods~\cite{prl_86_3_377-380_2001, prl_87_19_190402_2001}. To utilise these methods for an arbitrary value of $\theta$ it is necessary to set up a hydrodynamic formalism. This involves the definition of the condensate's density, $n$, phase, $S$, and superfluid velocity, $\mathbf{v}$, via the relations~\cite{pitaevskiistringaribec}:
\begin{align}
  \psi &= \sqrt{n}e^{iS}, \label{eq:nsdef} \\
  \mathbf{v} &= \nabla S. \label{eq:vdef}
\end{align}
Substituting Eqs.~\eqref{eq:nsdef} and \eqref{eq:vdef} into Eq.~\eqref{eq:rescaledgpe} yields a pair of hydrodynamic equations given by~\cite{advphys_57_6_539-616_2008, rmp_81_2_647-691_2009}:
\begin{align}
  \frac{\partial n}{\partial t} &= -\nabla\cdot\left[n\left(\mathbf{v}-\mathbf{\Omega}\times\mathbf{r}\right)\right], \label{eq:continuity} \\
  \frac{\partial\mathbf{v}}{\partial t} &= -\nabla\left\lbrace\frac{\mathbf{v}^2}{2} + V_{\text{T}} + \tilde{g}n - \mathbf{v}\cdot(\mathbf{\Omega}\times\mathbf{r}) - \frac{\nabla^2\left(\sqrt{n}\right)}{2\sqrt{n}}\right\rbrace. \label{eq:euler}
\end{align}

\begin{figure}[h]
\centering
\includegraphics[width=\linewidth]{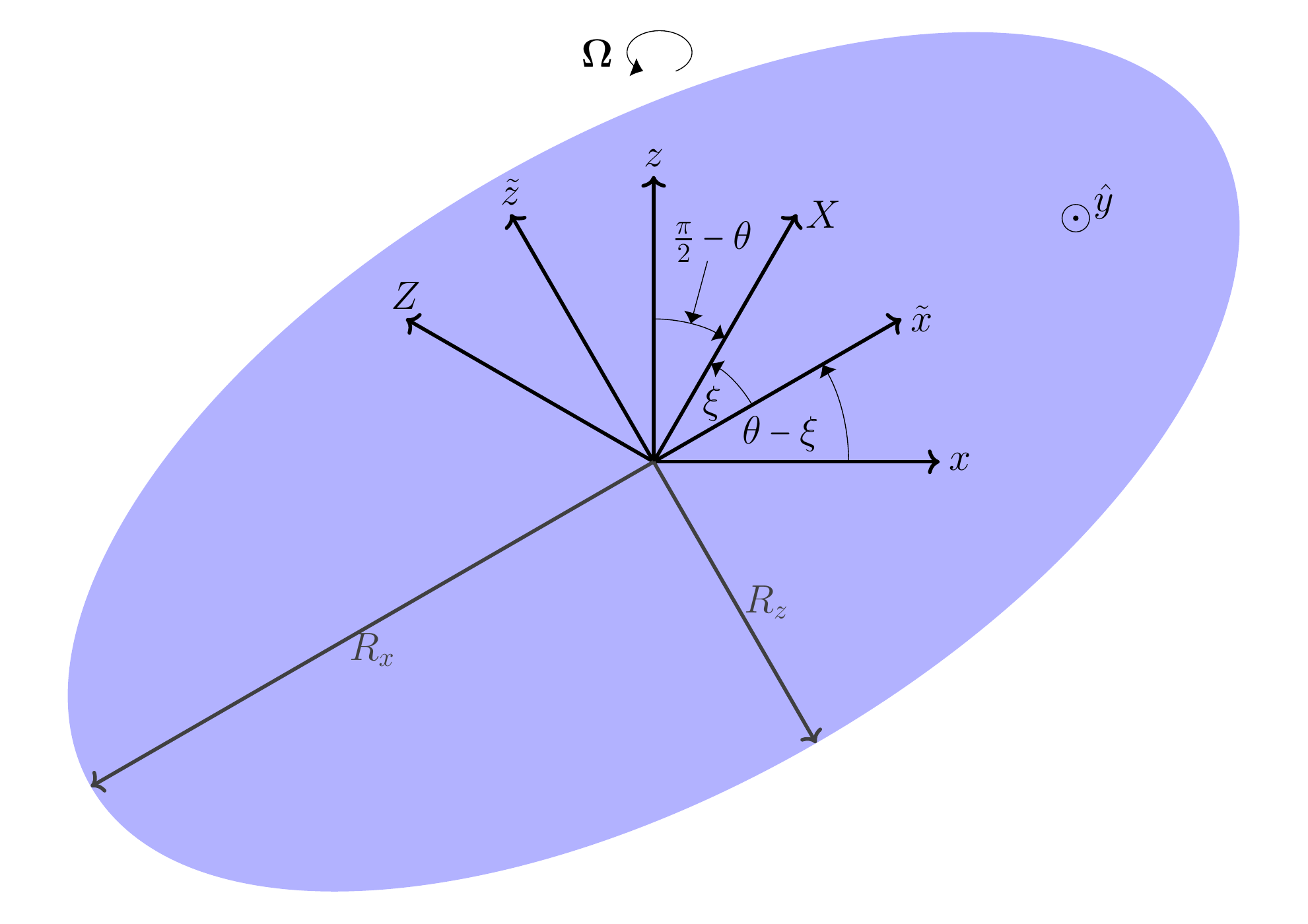}
\caption{Shaded cross-section, at $y = Y = \tilde{y} = 0$, of the ellipsoidal surface of constant density for a Thomas-Fermi stationary state with its semi-axes along the $\tilde{x}$- and $\tilde{z}$-axes, $R_x$ and $R_z$, respectively, illustrated for reference. The Cartesian axes corresponding to the coordinate frames $\mathbf{r}$, $\mathbf{R}$, and $\tilde{\mathbf{r}}$ are overlaid on the cross-section, and $\mathbf{\Omega} \parallel \hat{z}$.}
\label{referenceframes}
\end{figure}

When $Na_s \gg l_{\perp}$ the \emph{quantum pressure} term in Eq.~\eqref{eq:euler}, $\nabla\left[\nabla^2(\sqrt{n})/\sqrt{n}\right]$, is negligible due to the minimal effects of zero-point kinetic energy fluctuations in the condensate~\cite{pra_51_2_1382-1386_1995, prl_76_1_6-9_1996, pitaevskiistringaribec}. In the Thomas-Fermi (TF) limit, where this term may be neglected, Eq.~\eqref{eq:euler} is approximated by the simplified form
\begin{equation}
  \frac{\partial\mathbf{v}}{\partial t} = -\nabla\left\lbrace\frac{\mathbf{v}^2}{2} + V_{\text{T}} + \tilde{g}n -\mathbf{v}\cdot(\mathbf{\Omega}\times\mathbf{r})\right\rbrace. \label{eq:tfeuler}
\end{equation}
We also note that the vector $\mathbf{\Omega}\times\mathbf{r}$ lies in the $x$-$y$ plane whereas the principal axes of the trap are given by $\hat{X}$, $\hat{Y}$, and $\hat{Z}$, with $\hat{x}$ and $\hat{X}$ not coinciding with each other unless the trap is not tilted. The resulting competition between the trapping and rotating-frame transformation terms necessitates the introduction of a second angle, $\xi$, and a third co-rotating coordinate frame, $\tilde{\mathbf{r}}$, in order to find the axes of symmetry of the solutions of Eqs.~\eqref{eq:continuity} and \eqref{eq:tfeuler}. Let us define $\xi$ and $\tilde{\mathbf{r}}$ via the transformation
\begin{align}
\begin{pmatrix}
\tilde{x} \\
\tilde{y} \\
\tilde{z}
\end{pmatrix} &=
\begin{pmatrix}
\cos\xi & 0 & -\sin\xi \\
0 & 1 & 0 \\
\sin\xi & 0 & \cos\xi
\end{pmatrix}
\begin{pmatrix}
X \\
Y \\
Z
\end{pmatrix} \nonumber \\
&=
\begin{pmatrix}
\cos(\theta-\xi) & 0 & \sin(\theta-\xi) \\
0 & 1 & 0 \\
-\sin(\theta-\xi) & 0 & \cos(\theta-\xi)
\end{pmatrix}
\begin{pmatrix}
x \\
y \\
z
\end{pmatrix}. \label{eq:adjusttiltframe}
\end{align}
In this new reference frame, the trapping is given by
\begin{align}
  V_{\text{T}}\left(\tilde{\mathbf{r}}\right) &= \frac{1}{2}\left[(1-\varepsilon)\left(\tilde{x}\cos\xi + \tilde{z}\sin\xi\right)^2 + (1+\varepsilon)\tilde{y}^2\right] \nonumber \\
  &+ \frac{1}{2}\gamma^2\left(\tilde{x}\sin\xi - \tilde{z}\cos\xi\right)^2, \label{eq:tiltedtraptildecoord}
\end{align}
while the rotating-frame term, $\mathbf{\Omega}\times\mathbf{r}$ transforms to
\begin{equation}
\mathbf{\Omega}\times\tilde{\mathbf{r}} = \Omega\left[\cos(\theta-\xi)\left(-\tilde{y}\hat{\tilde{x}} + \tilde{x}\hat{\tilde{y}}\right) + \sin(\theta-\xi)\left(\tilde{y}\hat{\tilde{z}} - \tilde{z}\hat{\tilde{y}}\right)\right]. \label{eq:galileitiltedtrap}
\end{equation}
To clarify the relationship between the co-rotating reference frames, we overlay the coordinate axes of $\mathbf{r}$, $\mathbf{R}$, and $\tilde{\mathbf{r}}$ at constant $y = Y = \tilde{y} = 0$ on a typical cross-section of an TF surface of constant density in Fig. \ref{referenceframes}.

\section{\label{sec:level2}Thomas-Fermi Stationary Solutions}
The stationary solutions of the GPE are specified through the condensate's chemical potential, $\mu$, via~\cite{pitaevskiistringaribec}
\begin{equation}
  \psi(\tilde{\mathbf{r}}, t) = \psi(\tilde{\mathbf{r}}, t = 0)\exp(-i\mu t).
\end{equation}
Therefore, the stationary state density, $n_{\text{TF}}$, and velocity, $\mathbf{v}_{\text{TF}}$, obey
\begin{gather}
  0 = \nabla\cdot\left[n\left(\mathbf{v}-\mathbf{\Omega}\times\tilde{\mathbf{r}}\right)\right], \label{eq:contstat} \\
  \nabla\mu = \nabla\left\lbrace\frac{\mathbf{v}_{\text{TF}}^2}{2} + V_{\text{T}} + \tilde{g}n_{\text{TF}} -\mathbf{v}_{\text{TF}}\cdot(\mathbf{\Omega}\times\tilde{\mathbf{r}})\right\rbrace. \label{eq:tfeulerstat}
\end{gather}
Let us impose the following \emph{Ans{\"a}tze} for $n_{\text{TF}}$ and $\mathbf{v}_{\text{TF}}$:
\begin{align}
  n_{\text{TF}}(\tilde{\mathbf{r}}) &= n_0\left(1 - \mathlarger{\sum_{i\in{x,y,z}}}\frac{\tilde{r}_i^2}{R_i^2}\right)\Theta\left(1 - \mathlarger{\sum_{i\in{x,y,z}}}\frac{\tilde{r}_i^2}{R_i^2}\right), \label{eq:nstat} \\
  \mathbf{v}_{\text{TF}}(\tilde{\mathbf{r}}) &= \nabla\left[\alpha_{xy}\tilde{x}\tilde{y} + \alpha_{yz}\tilde{y}\tilde{z} + \alpha_{zx}\tilde{z}\tilde{x}\right]. \label{eq:modifyrecati}
\end{align}
Here, $n_0 = 15/(8\pi R_xR_yR_z)$ is a normalization parameter that ensures that $n_{\text{TF}}$ obeys Eq.~\eqref{eq:normalization}~\cite{pra_51_2_1382-1386_1995, prl_76_1_6-9_1996}. The form of Eq.~\eqref{eq:nstat} shows that the angle $\xi$ in the coordinate transformation given by Eq.~\eqref{eq:adjusttiltframe} is fixed by the requirement that the principal axes of the TF stationary state density coincide with the Cartesian axes of the $\mathbf{r}$ coordinate frame. The parameters $\lbrace R_i\rbrace$ thus denote the semi-axes of the paraboloid TF profile along the $\tilde{r}_i$-axis. We illustrate these features in the TF density cross-section in Fig.~\ref{referenceframes} by labeling the ellipsoid's semi-axes along $\hat{\tilde{x}}$ and $\hat{\tilde{z}}$ as $R_x$ and $R_z$, respectively.

Equation~\eqref{eq:modifyrecati} is consistent with the quadrupolar flow of a TF stationary state in an untilted harmonic trap ($\theta = \xi = 0$) rotating about the $z$-axis, $\mathbf{v}_{\text{TF}} = \alpha\nabla(xy)$~\cite{prl_86_3_377-380_2001}. An inspection of Eq.~\eqref{eq:contstat} shows that the $k$th component of $\mathbf{v}$, $\sum_{j\neq k}\alpha_{jk}\tilde{r}_j$, is nonzero only if $\epsilon_{ijk}\Omega_i\tilde{r}_j \neq 0$, which in turn shows that $\alpha_{ij} \neq 0$ only if $\epsilon_{ijk}\Omega_k \neq 0$. This suggests that for the problem at hand, we have $\alpha_{zx} = 0$ since $\Omega_y = 0$. By substituting Eq.~\eqref{eq:nstat} and~\eqref{eq:modifyrecati} into Eq.~\eqref{eq:contstat} and equating the coefficients of the spatial coordinates, we can verify the property that $\alpha_{zx}$ is null and also derive the relations
\begin{align}
\alpha \equiv \alpha_{xy} &= \left(\frac{\kappa_x^2 - \kappa_y^2}{\kappa_x^2 + \kappa_y^2}\right)\Omega\cos(\theta-\xi), \label{eq:alphadefn} \\
\delta \equiv \alpha_{zx} &= \left(\frac{\kappa_y^2 - 1}{\kappa_y^2 + 1}\right)\Omega\sin(\theta-\xi), \label{eq:deltadefn}
\end{align}
where $\kappa_x = R_x/R_z$ and $\kappa_y = R_y/R_z$. Thus the trial solution employed for the velocity field is
\begin{equation}
  \mathbf{v}_{\text{TF}}(\tilde{\mathbf{r}}) = \alpha\nabla(\tilde{x}\tilde{y}) + \delta\nabla(\tilde{y}\tilde{z}). \label{eq:vstat}
\end{equation}
This quadrupolar profile for the velocity field, and thereby the spatial dependence of the condensate's phase, may be considered as the quantum analog of the classical velocity potential for an inviscid fluid inside an ellipsoid container rotating about a non-principal axis of the ellipsoid~\cite{lambhydrodynamics, landaulifshitzvol6fluidmechanics}. In both systems, solid-body rotation is possible only when the density is asymmetric about the rotation axis. We also note that Eqs.~\eqref{eq:alphadefn} and \eqref{eq:deltadefn} are formally similar to the equations of motion appearing in the context of the rotational energy bands in the tilted-axis cranked shell model of rotating triaxial nuclei~\cite{prc_65_5_054304_2002}.

The problem of determining the stationary solutions of Eqs.~\eqref{eq:contstat} and \eqref{eq:tfeulerstat} may now be reduced to solving a set of five self-consistency relations for $\lbrace \kappa_x, \kappa_y, \alpha, \delta, \xi\rbrace$. These are obtained by substituting Eqs.~\eqref{eq:nstat} and \eqref{eq:vstat} into Eq.~\eqref{eq:tfeulerstat} and subsequently reading off the coefficients of like terms. Firstly, from the coefficients of $\tilde{x}^2$, $\tilde{y}^2$ and $\tilde{z}^2$, we find that the TF semi-axes are given by
\begin{equation}
  R_i^2 = \frac{2\tilde{g}n_0}{\tilde{\omega}_i^2}. \label{eq:tfradii}
\end{equation}
In Eq.~\eqref{eq:tfradii}, we make use of generalized harmonic trapping frequencies, $\tilde{\omega}_i^2$, that are defined as
\begin{align}
  \tilde{\omega}_x^2 &= (1-\varepsilon)\cos^2\xi + \gamma^2\sin^2\xi + \alpha^2 - 2\Omega\alpha\cos(\theta - \xi), \label{eq:omegaxeff} \\
  \tilde{\omega}_y^2 &= 1+\varepsilon + \alpha^2 + \delta^2 + 2\Omega[\alpha\cos(\theta-\xi)-\delta\sin(\theta-\xi)], \label{eq:omegayeff} \\
  \tilde{\omega}_z^2 &= \gamma^2\cos^2\xi + (1-\varepsilon)\sin^2\xi + \delta^2 + 2\Omega\delta\sin(\theta-\xi). \label{eq:omegazeff}
\end{align}
This implies that the quantities $\kappa_x$ and $\kappa_y$ obey
\begin{equation}
  \kappa_i^2 = \frac{\tilde{\omega}_z^2}{\tilde{\omega}_i^2}\,:\,i = x,y. \label{eq:tfratiosols}
\end{equation}
By recognizing that there is no $\tilde{x}\tilde{z}$ term in Eq.~\eqref{eq:nstat}, we also obtain the condition that
\begin{equation}
  (1 - \varepsilon - \gamma^2)\sin\xi\cos\xi + \alpha\delta + \Omega[\alpha\sin(\theta - \xi) - \delta\cos(\theta - \xi)] = 0. \label{eq:xzcoeffzero}
\end{equation}
The two final self-consistency relations are obtained via substituting Eq.~\eqref{eq:tfradii} into Eqs.~\eqref{eq:alphadefn} and \eqref{eq:deltadefn}, which yields
\begin{align}
  [\alpha+\Omega\cos(\theta-\xi)]\tilde{\omega}_x^2 &+ [\alpha-\Omega\cos(\theta-\xi)]\tilde{\omega}_y^2 = 0, \label{eq:alphaeqn} \\
  [\delta+\Omega\sin(\theta-\xi)]\tilde{\omega}_y^2 &+ [\delta-\Omega\sin(\theta-\xi)]\tilde{\omega}_z^2 = 0. \label{eq:deltaeqn}
\end{align}

Equations \eqref{eq:tfratiosols} --~\eqref{eq:deltaeqn} describe branches of stationary solutions as functions of $\Omega$ that terminate when one or more of $\tilde{\omega}_x, \tilde{\omega}_y, \tilde{\omega}_z$ equal zero. The locations of these endpoints determine the number of real stationary solutions for a given value of $\Omega$. We identify four such limits which are of use to us, noting that a rotation of the rotating frame by $\pi/2$ about the $\tilde{y}$-axis transforms $\tilde{x}$ to $\tilde{z}$:
\begin{enumerate}[label=\alph*)]
  \item \begin{center}$\tilde{\omega}_x \rightarrow 0$ and $\tilde{\omega}_y,\text{}\tilde{\omega}_z \neq 0$,\end{center} \label{en:endpointx}
  \item \begin{center}$\tilde{\omega}_y \rightarrow 0$ and $\tilde{\omega}_x,\text{}\tilde{\omega}_z \neq 0$,\end{center} \label{en:endpointy}
  \item \begin{center}$\tilde{\omega}_x,\text{}\tilde{\omega}_y \rightarrow 0$ and $\tilde{\omega}_z \neq 0$,\end{center} \label{en:endpointxy}
  \item \begin{center}$\tilde{\omega}_y,\text{}\tilde{\omega}_z \rightarrow 0$ and $\tilde{\omega}_x \neq 0$.\end{center} \label{en:endpointyz}
\end{enumerate}
For the remainder of this paper, the subscripts $xc$, $yc$, $xyc$, and $yzc$ are used to denote the values of quantities such as $\xi$ in the limits~\ref{en:endpointx}, \ref{en:endpointy}, \ref{en:endpointxy} and \ref{en:endpointyz}, respectively. A detailed description of the self-consistency relations satisfied by $\Omega$, $\alpha$, $\delta$ and $\xi$ at each of these limits is provided in Appendix~\ref{sec:level7}. We also provide a description of how the shape of the TF distribution can be understood via inspection of the signs of $\alpha$, $\delta$ and $\theta - \xi$ can be found in Appendix~\ref{sec:level8}.

\section{\label{sec:level3}Stationary Solution Branches}
Keeping in mind the possible limits of the stationary solution branches, $\Omega \rightarrow \lbrace\Omega_{xc}, \Omega_{yc}, \Omega_{xyc}, \Omega_{yzc}\rbrace$, we proceed to solve Eqs.~\eqref{eq:tfratiosols} -- \eqref{eq:deltaeqn} and plot the resulting values of $\alpha$, $\delta$, and $\xi$ as functions of $\Omega$ for fixed values of $\theta$, $\gamma$ and $\varepsilon$. To provide a representative sample of the variety of trapping regimes, we analyze the following cases:
\begin{enumerate}
\item $\gamma = 3/4$, $\varepsilon = 0$, $\theta \in \lbrace 0, \pi/8, \pi/4, 3\pi/8\rbrace$,
\item $\gamma = 4/3$, $\varepsilon = 0.05$, $\theta \in \lbrace 0, \pi/8, \pi/4, 3\pi/8\rbrace$.
\end{enumerate}
When $\Omega = 0$, stationary states in the harmonic trap described by case $1$ are prolate and are axially symmetric about $\hat{z}$, while those for case $2$ are oblate and do not exhibit this axial symmetry. We do not analyze the trap tilting angle $\theta = \pi/2$ as this limit is easily transformed to an untilted trap with a different set of trapping frequencies by rotating the coordinate frame about $\hat{y}$ by $\pi/2$.

\subsection{\label{sec:level3.1}Prolate, Symmetric Trapping}
We initially focus on the prolate, symmetric trap, where $\gamma = 3/4$ and $\varepsilon = 0$. For this trap, we specify the rotational frequencies as defined by cases a) -- d) in Sec.~\ref{sec:level2} in Table~\ref{tab:omcrit1}:

\begin{table}[h]
\caption{Endpoints of the branches - $\gamma = 3/4$, $\varepsilon = 0$ \label{tab:omcrit1}}
\begin{ruledtabular}
\begin{tabular}{c||c|c|c|c}
 & $\Omega_{xc}/\omega_{\perp}$ & $\Omega_{yc}/\omega_{\perp}$ & $\Omega_{xyc}/\omega_{\perp}$ & $\Omega_{yzc}/\omega_{\perp}$ \\  \hline
$\theta = 0$ & 1 & 1 & 1.75 & 1.75 \\
$\theta = \pi/8$ & 0.9475 & 1 & 1.8958 & 1.6495 \\
$\theta = \pi/4$ & 0.8485 & 1 & 1.9899 & 1.6578 \\
$\theta = 3\pi/8$ & 0.7752 & 1 & 1.9833 & 1.7563 \\
\end{tabular}
\end{ruledtabular}
\end{table}

\begin{figure}[h]
\includegraphics[width=\linewidth]{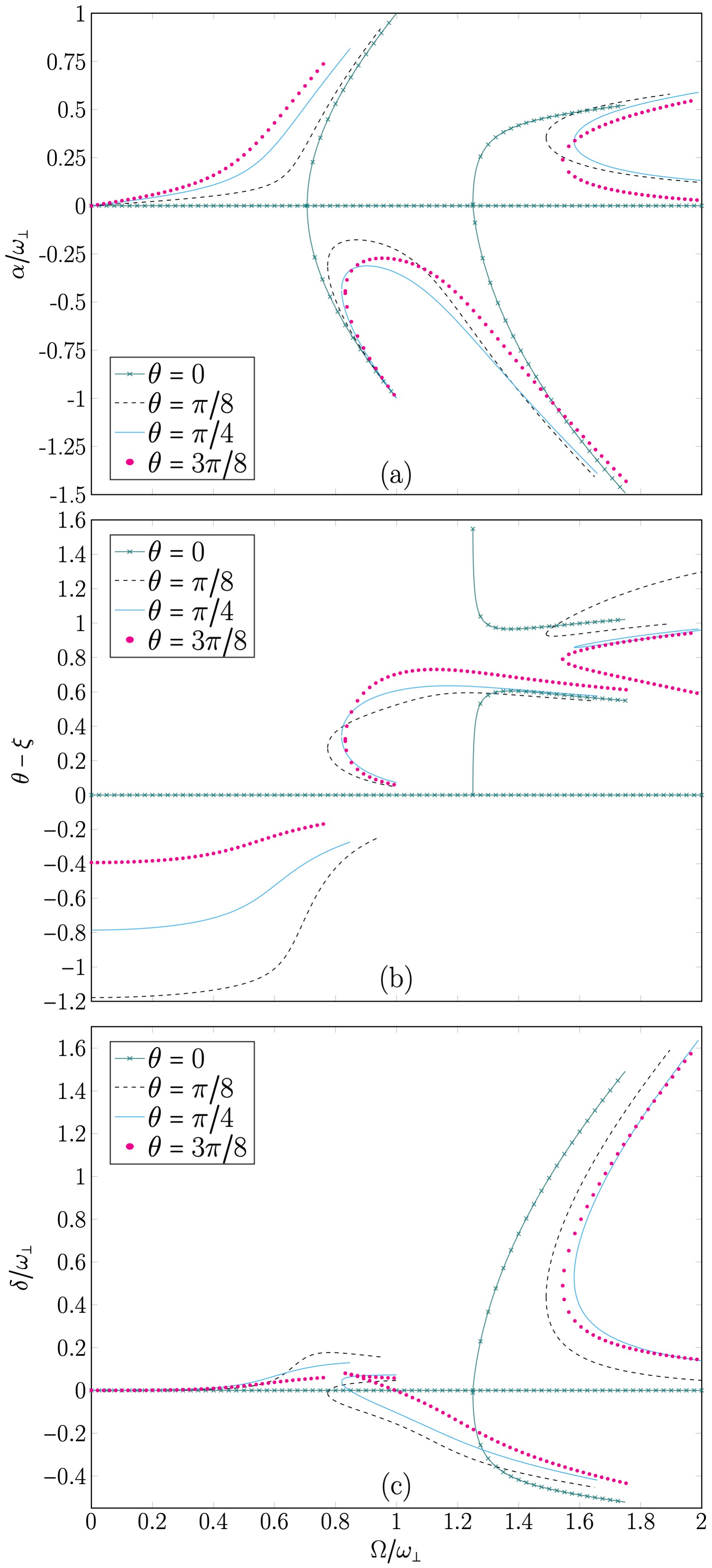}
\vspace*{-5mm}
\caption{Stationary solutions as a function of $\Omega$ for $\alpha$ (a), $\theta - \xi$ (b), and $\delta$ (c), when $\gamma = 3/4$, $\varepsilon = 0$, $\theta \in \lbrace 0, \pi/8, \pi/4, 3\pi/8\rbrace$.}
\label{epsilon0gamma0point75thetaalltf}
\end{figure}

In Fig.~\ref{epsilon0gamma0point75thetaalltf}(a), we plot $\alpha$ as a function of $\Omega$ for the values of $\theta$ listed in Table~\ref{tab:omcrit1}. Here we see that there exist five distinct stationary solution branches, four of which exhibit the endpoints defined by cases a) -- d). Initially we describe the limit $\theta = 0$, as explored in previous theoretical studies. For a trap with axial symmetry about the rotation axis, i.e. $\varepsilon = 0$, an $\alpha = 0$ stationary solution exists for all $\Omega \geq 0$, while two further solutions emerge at the rotational bifurcation frequency $\Omega = \Omega_{\text{b}1} \equiv \omega_{\perp}/\sqrt{2}$~\cite{prl_86_3_377-380_2001, pra_73_6_061603r_2006, jphysb_40_18_3615-3628_2007}. We note that the position of this bifurcation is attributable to an energetic instability of the $l = 2,\,m = 2$ quadrupolar surface mode, which has a frequency $\omega(l = 2,\,m = 2) = \sqrt{2}\omega_{\perp} - 2\Omega$ and is thus energetically favourable for $\Omega \geq \Omega_{\text{b}1}$~\cite{prl_86_3_377-380_2001}. These additional stationary solutions are symmetric about the $\Omega$ axis and terminate in the limit $\Omega \rightarrow \Omega_{xc} = \Omega_{yc} = \omega_{\perp}$, where $\alpha \rightarrow \omega_{\perp}$ as well. Furthermore, we find evidence for the existence of a second bifurcation where two more stationary solutions emerge from the stationary solution defined by $\alpha = 0$ and terminate when $\Omega \rightarrow \Omega_{xyc} = \Omega_{yzc} = 1.75\omega_{\perp}$. We attribute the existence of this bifurcation to the energetic instability of the $l = 2,\,m = 1$ quadrupole mode, which boasts the frequency $\omega(l = 2,\,m = 1) = \omega_{\perp}\sqrt{1 + \gamma^2} - \Omega$~\cite{pitaevskiistringaribec} and is therefore associated with the bifurcation frequency $\Omega_{\text{b}2} = 5\omega_{\perp}/4$ when $\gamma = 3/4$. These new branches are not symmetric about the $\Omega$ axis unlike those emerging from the $m = 1$ bifurcation, and their existence had not previously been predicted in the context of rotating BECs due to the omission of the additional degrees of freedom given by $\delta$ and $\xi$. However, we note that the energetic instability of an $l = 2,\,m = 1$ surface mode causes similar bifurcations in other systems. For instance, in a rotating reference frame, the equilibrium density of a irrotational gravitationally-bound fluid can undergo just such a bifurcation from a Maclaurin spheroid to a tilted Riemann ellipsoid~\cite{physfluids_8_12_3414-3422_1996}.

The new class of stationary solutions described by $\theta \neq 0$ behaves markedly differently to those for the untilted trap. When $\Omega = 0$ we have a solution defined by $\alpha = 0$ and this solution, which we denote as Branch I, persists for $\Omega < \min\lbrace\Omega_{xc}, \Omega_{yc}\rbrace$. From Table~\ref{tab:omcrit1}, this rotation frequency is given by $\Omega = \Omega_{xc}$ for all of the values of $\theta$ that we consider in this case. Branch I is, in general, the solution that the condensate will follow in response to a quasi-adiabatic acceleration of the trap's rotation frequency from zero. Two additional, connected, branches emerge at a bifurcation frequency, denoted as $\Omega_{\text{b}1}$, and initially have values of $\alpha$ with opposite sign to the first solution. One of these solutions, denoted here as Branch II, terminates at $\Omega = \max\lbrace\Omega_{xc}, \Omega_{yc}\rbrace \equiv \omega_{\perp}$. The other solution, denoted here as Branch III, persists until the endpoint defined by $\Omega = \min\lbrace\Omega_{xyc}, \Omega_{yzc}\rbrace$, which is equivalent to $\Omega = \Omega_{yzc}$ for this trap. The behavior of Branch III contrasts with that of the solutions for $\theta = 0$, where it is possible for a condensate to follow the same solution branch from $\Omega = \Omega_{\text{b}1}$ till $\Omega \rightarrow \infty$. A second bifurcation frequency, $\Omega_{\text{b}2}$, heralds the emergence of an additional pair of connected branches that exhibit the same sign of $\alpha$. One of these, denoted here as Branch IV, terminates when $\Omega \rightarrow \max\lbrace\Omega_{xyc}, \Omega_{yzc}\rbrace \equiv \Omega_{xyc}$, while the other solution, denoted here as Branch V, exists for $\Omega \in [\Omega_{\text{b}2}, +\infty)$ and is the only solution that exists for $\Omega > \max\lbrace\Omega_{xyc}, \Omega_{yzc}\rbrace$.

We also present the corresponding solutions of $\theta - \xi$, as a function of $\Omega$, in Fig.~\ref{epsilon0gamma0point75thetaalltf}(b) where we observe that both of the bifurcations are clearly evident in the behavior of $\theta - \xi$ as well as that of $\alpha$. Furthermore, for $\theta = 0$, the solutions that emerge at $\Omega = \Omega_{\text{b}2}$ and terminate at $\Omega = \Omega_{xyc} = \Omega_{yzc} = 1.75\omega_{\perp}$ are closely related to each other; they correspond to density profiles with the identical TF semi-axes but with opposite tilting angles about the rotation axis. As such, their respective values of $\xi$ are symmetric about the value $\xi = -\pi/4$. In Fig.~\ref{epsilon0gamma0point75thetaalltf}(c), where $\delta$ is plotted as a function of $\Omega$, we find that for the $\theta = 0$ branches emerging when $\Omega = \Omega_{\text{b}2}$, the values of $\alpha$ for one branch are equivalent to those of $-\delta$ for the other branch. We also note that unlike the corresponding behavior of $\alpha$ and $\xi$, a qualitative discrepancy in $\delta$ along Branch I for $\theta = \pi/8$ is evident when compared to the angles $\theta = \pi/4$ and $\theta = 3\pi/8$. Specifically, $\delta$ is a monotonically increasing function of $\Omega$ when $\theta = \pi/4$ and $\theta = 3\pi/8$ but exhibits a maximum at $\Omega\approx 0.78\omega_{\perp}$ when $\theta = \pi/8$. However, such qualitative differences with respect to the trap tilting angle are not exhibited by Branches II - V.

\subsection{\label{sec:level3.2}Oblate, Asymmetric Trapping}
We proceed to discuss the condensate's behavior in the oblate, asymmetric trap where $\gamma = 4/3$ and $\varepsilon = 0.05$. Here, the lack of axial symmetry of the trapping along any axis results in the features of the stationary solutions being qualitatively different to those described in Sec.~\ref{sec:level3.1}. In Table~\ref{tab:omcrit2}, we specify the rotation frequencies that correspond to the termination cases a) -- d):

\begin{table}[h]
\caption{Endpoints of the branches - case $2$ \label{tab:omcrit2}}
\begin{ruledtabular}
\begin{tabular}{c||c|c|c|c}
 & $\Omega_{xc}/\omega_{\perp}$ & $\Omega_{yc}/\omega_{\perp}$ & $\Omega_{xyc}/\omega_{\perp}$ & $\Omega_{yzc}/\omega_{\perp}$ \\  \hline
$\theta = 0$ & 0.9747 & 1.0247 & 2.3334 & 2.3334 \\
$\theta = \pi/8$ & 1.0097 & 1.0247 & 2.1311 & 2.4914 \\
$\theta = \pi/4$ & 1.1128 & 1.0247 & 1.9924 & 2.5176 \\
$\theta = 3\pi/8$ & 1.2556 & 1.0247 & 2.0012 & 2.3892 \\
\end{tabular}
\end{ruledtabular}
\end{table}

\begin{figure}[h]
\includegraphics[width=\linewidth]{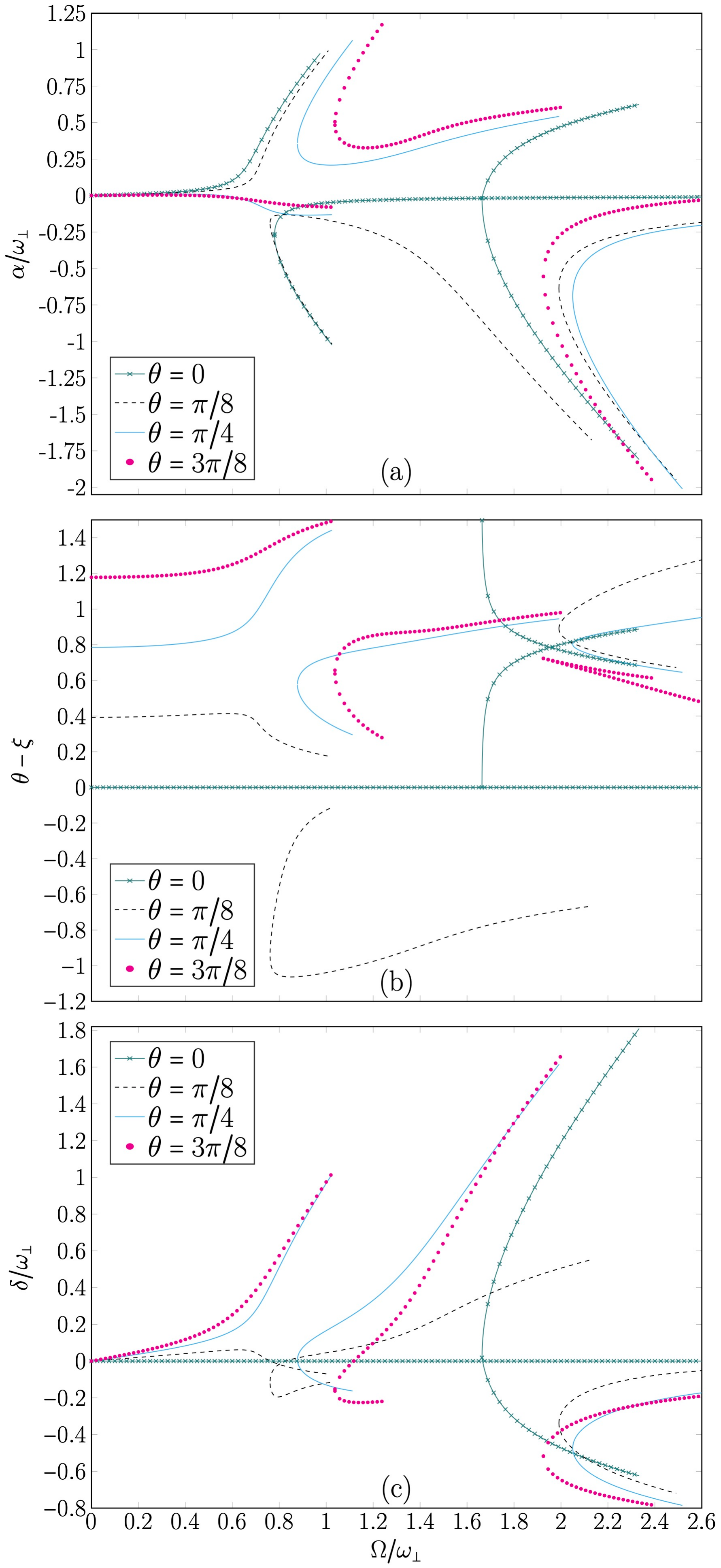}
\vspace*{-5mm}
\caption{Stationary solutions as a function of $\Omega$ for $\alpha$ (a), $\theta - \xi$ (b), and $\delta$ (c), when $\gamma = 4/3$, $\varepsilon = 0.05$, $\theta \in \lbrace 0, \pi/8, \pi/4, 3\pi/8\rbrace$.}
\label{epsilon0point05gamma4over3thetaalltf}
\end{figure}

When the rotating trap is untilted, i.e. $\theta = 0$, the stationary solutions corresponding to Branches I, II and III are also untilted, i.e. $\theta = \xi = 0$. We find that Branch I, for which $\alpha \geq 0$, terminates when $\Omega = \Omega_{xc} = \omega_{\perp}\sqrt{1-\varepsilon}$. Branches II and III, which both exhibit $\alpha < 0$, are connected at the bifurcation frequency $\Omega = \Omega_{\text{b}1}$ but are disconnected from Branch I. While Branch II terminates when $\Omega = \Omega_{yc} = \omega_{\perp}\sqrt{1+\varepsilon}$, Branch III is characterized by $\alpha$ monotonically tending to zero as $\Omega \rightarrow \infty$~\cite{prl_86_3_377-380_2001, pra_73_6_061603r_2006, jphysb_40_18_3615-3628_2007}. The extra degrees of freedom that are represented by $\delta$ and $\xi$ manifest themselves when $\theta = 0$ through the presence of the additional, previously unknown, branches IV and V, which are connected at $\Omega = \Omega_{\text{b2}}$ and terminate at the same rotation frequency, $\Omega = \Omega_{xyc} = \Omega_{yzc}$. However, when the rotating trap is tilted the stationary solutions more closely resemble those in Sec.~\ref{sec:level3.1} except that $\Omega_{xc} < \Omega_{yc} = \omega_{\perp}\sqrt{1+\varepsilon}$ when $\theta = 0, \pi/8$ and $\Omega_{xc} > \omega_{\perp}\sqrt{1+\varepsilon}$ when $\theta \in \pi/4, 3\pi/8$; the crossover, where $\Omega_{xc} = \Omega_{yc}$, occurs when $\theta \approx 0.4693 \equiv 26.89^{\circ}$. This results in Branches I -- III possessing the opposite signs for $\Omega_{xc} > \Omega_{yc}$ when $\theta \in \pi/4, 3\pi/8$ to the solutions when $\Omega_{xc} < \Omega_{yc}$, a feature not seen in Sec.~\ref{sec:level3.1}. This behavior is demonstrated in Fig.~\ref{epsilon0point05gamma4over3thetaalltf}(a), where we have plotted $\alpha$ as a function of $\Omega$ for the angles $\theta \in \lbrace 0, \pi/8, \pi/4, 3\pi/8\rbrace$.

We have also plotted $\theta - \xi$ as a function of $\Omega$ in Fig.~ \ref{epsilon0point05gamma4over3thetaalltf}(b) for this oblate, axially asymmetric trap, and thereby find that $\theta - \xi$ similarly behaves differently for Branch I when $\theta = \pi/8$ as compared to the angles $\theta = \pi/4$ and $\theta = 3\pi/8$. For instance, the behavior of Branch I for $\theta = \pi/8$ is not monotonic but has a maximum at $\Omega \approx 0.65\omega_{\perp}$. This contrasts sharply with the monotonic behavior of $\theta - \xi$ as a function of $\Omega$ for $\theta = \pi/4$ and $\theta = 3\pi/8$. However, Branches II - V exhibit merely quantitative differences with respect to the tilting angle. As in Sec.~\ref{sec:level3.1}, the values of $\xi$ for the branches that emerge when $\Omega = \Omega_{\text{b}2}$ are symmetric about the value $\xi = -\pi/4$, suggesting that the density profiles for these two branches are physically equivalent with the same TF semi-axes but exhibit opposite tilting angles about the rotation axis. Thus the values of $\alpha$ for one branch is equivalent to those of $-\delta$ for the other branch, which may be inferred from the corresponding plots of $\delta$, as a function of $\Omega$, that are provided in Fig.~ \ref{epsilon0point05gamma4over3thetaalltf}(c). Interestingly, the maximum of $\theta - \xi$ for $\theta = \pi/8$ along Branch I when $\Omega \approx 0.65\omega_{\perp}$ is reflected in a similar maximum in $\delta$, which eventually attains negative values as $\Omega \rightarrow \Omega_{xc}$.

\section{\label{sec:level4}Linearized Time-Dependent Hydrodynamics}
Via the hydrodynamic formalism elucidated in Secs.~\ref{sec:level2} and \ref{sec:level3}, we have shown in Sec.~\ref{sec:level3} that the tilting of a rotating harmonic trap induces a nontrivial tilting angle of the condensate's TF stationary state density. The hydrodynamic formalism may also be used to determine the parametric domain of dynamical stability against environmental perturbations, a procedure that has been achieved in the $\theta = 0$ limit~\cite{prl_87_19_190402_2001}. Let us specifically address the scenario where the rotation frequency, $\Omega$, is quasi-adiabatically accelerated from zero for a fixed choice of $\varepsilon$, $\gamma$, and $\theta$. In the TF limit, the condensate will follow the stationary solution Branch I and therefore we solely investigate the dynamical stability of Branch I.

In general, the perturbating of a trapped BEC in a stationary state can excite one or more of its collective modes. For perturbations of sufficiently small magnitude the condensate's response may be assumed to be linear and the collective excitations may be obtained by linearizing Eqs.~\eqref{eq:continuity} and \eqref{eq:tfeuler} about the TF stationary state. In this formalism the collective modes are expressed as time-dependent fluctuations of the density and phase that are equivalent to linear combinations of the solutions of the Bogoliubov--de Gennes equations~\cite{pra_54_5_4204-4212_1996, pra_58_4_3168-3179_1998, pitaevskiistringaribec}. To determine the spectrum of collective modes, we write:
\begin{align}
n(\tilde{\mathbf{r}}, t) &= n_{\text{TF}}(\tilde{\mathbf{r}}) + \delta n(\tilde{\mathbf{r}}, t), \label{eq:densitypert} \\
S(\tilde{\mathbf{r}}, t) &= S_{\text{TF}}(\tilde{\mathbf{r}}, t) + \delta S(\tilde{\mathbf{r}}, t). \label{eq:phasepert}
\end{align}
Here, $S_{\text{TF}}(\tilde{\mathbf{r}}, t) = -\mu t + \alpha\tilde{x}\tilde{y} + \delta\tilde{y}\tilde{z}$. The subsequent linearization of Eqs.~\eqref{eq:continuity} and \eqref{eq:tfeuler} is equivalent to neglecting contributions from terms that are quadratic in the density and phase fluctuations, $\delta n$ and $\delta S$, respectively. This results in a coupled set of first-order equations for the time evolution of the fluctuations that is given by~\cite{prl_87_19_190402_2001, pra_73_6_061603r_2006}
\begin{gather}
\frac{\partial}{\partial t}
\begin{pmatrix}
\delta S \\
\delta n
\end{pmatrix}
=
\mathcal{M}
\begin{pmatrix}
\delta S \\
\delta n
\end{pmatrix}, \label{eq:perteqns} \\
\mathcal{M} = -\begin{pmatrix}
\mathbf{v}_c\cdot\nabla & \tilde{g} \\
\nabla\cdot\left(n_{\text{TF}}\nabla\right) & \mathbf{v}_c\cdot\nabla
\end{pmatrix}, \label{eq:pertmatrix} \\
\mathbf{v}_c = \nabla S_{\text{TF}} - \mathbf{\Omega}\times\tilde{\mathbf{r}}. \label{eq:labvel}
\end{gather}
Hence we can express each collective mode, indexed by $\nu$, as a combination of a density fluctuation, $\delta n_{\nu}(\tilde{\mathbf{r}})e^{\lambda_{\nu}t}$, and a phase fluctuation, $\delta S_{\nu}(\tilde{\mathbf{r}})e^{\lambda_{\nu}t}$, that satisfies Eq.~\eqref{eq:perteqns} if the constant $\lambda_{\nu}$ is an eigenvalue of the operator $\mathcal{M}$.

Since the time-dependence of the collective modes is exponential it is evident that, to linear order, the dynamical stability of a stationary state is determined by the set of all eigenvalues of $\mathcal{M}$, $\lbrace\lambda_{\nu}\rbrace$. If a given eigenvalue has a positive real component, the amplitude of the corresponding collective mode will grow exponentially in time and will overwhelm the stationary state, rendering the stationary state dynamically unstable. Conversely we have dynamical stability only if all of the eigenvalues of $\mathcal{M}$ have a negative real component, while purely imaginary eigenvalues are characteristic of excitations with an infinite lifetime. To diagonalise $\mathcal{M}$, we expand $\delta n$ and $\delta S)$ as polynomials in $\mathbb{R}^3$~\cite{prl_87_19_190402_2001, pra_73_6_061603r_2006}. Since it is not possible to consider all possible collective modes, we truncate the polynomial expansion of the fluctuations such that the maximum allowed order of the polynomials is $N_{\text{max}} = 10$. This proves to be a sufficiently high order to explore the dynamical stability of the stationary states in the linearized regime. However, we note that even if no unstable modes are found from this procedure for a given stationary state, it is not a guarantee of dynamical stability as a higher value of $N_{\text{max}}$ may admit a collective mode whose eigenvalue has a positive real component. Furthermore, by limiting our analysis to the linearized regime, we neglect nonlinear effects that could destabilize modes that are stable at linear order in the fluctuations.

\begin{figure}[h]
\includegraphics[width=\linewidth]{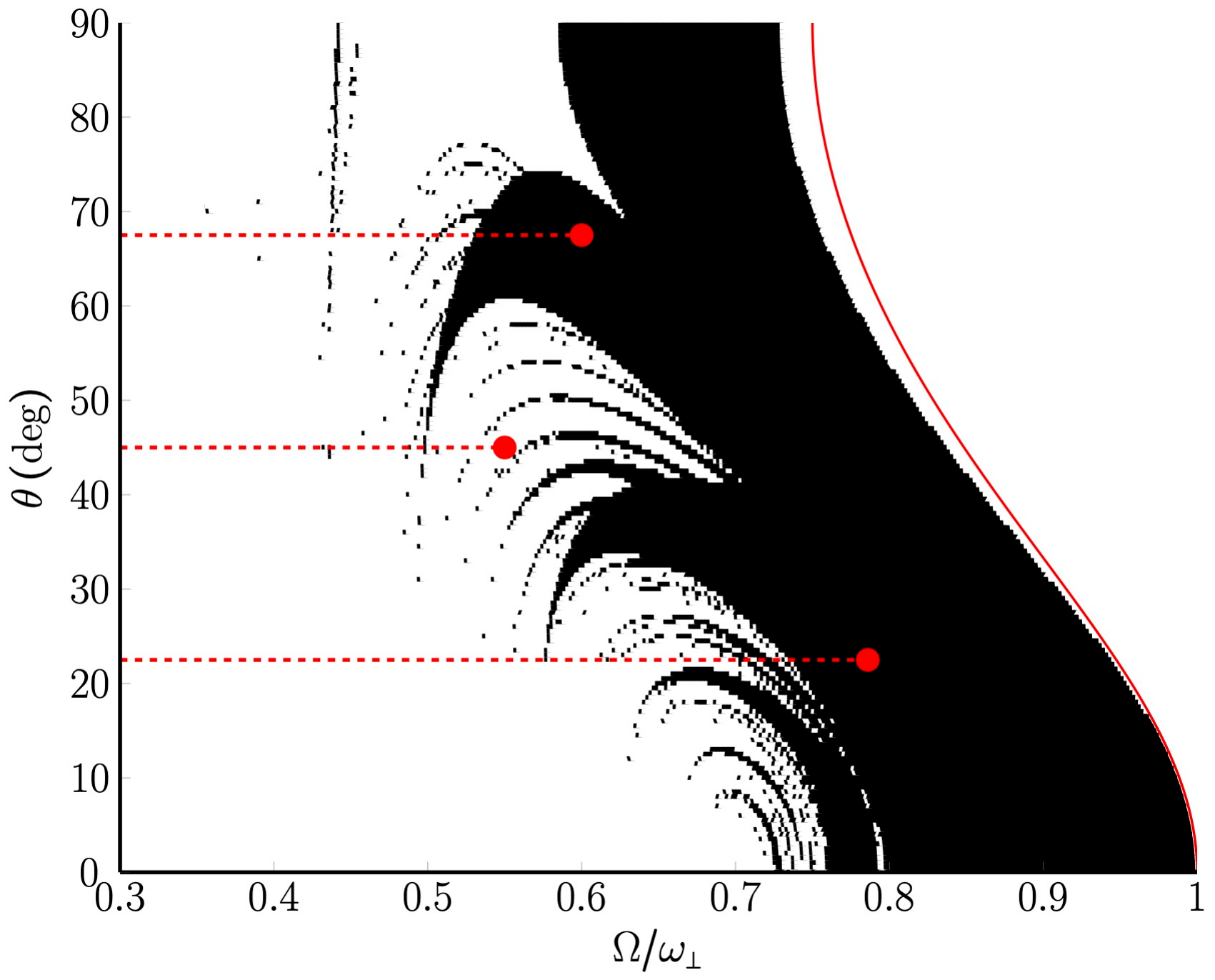}
\vspace*{-5mm}
\caption{Phase diagram of the dynamical stability of Branch I for $\varepsilon = 0,\,\gamma = 3/4$, with $N_{\text{max}} = 10$; Branch I is dynamically unstable at the shaded points of parameter space. The red dashed lines and markers denote the trajectories of the GPE simulations and the corresponding instability frequency, respectively. The red unbroken curve denotes the endpoints of Branch I, $\Omega = \min\lbrace\Omega_{xc},\Omega_{yc}\rbrace$.}
\label{epsilon0gamma0point75upperheaviside}
\end{figure}

\begin{figure}[h]
\includegraphics[width=\linewidth]{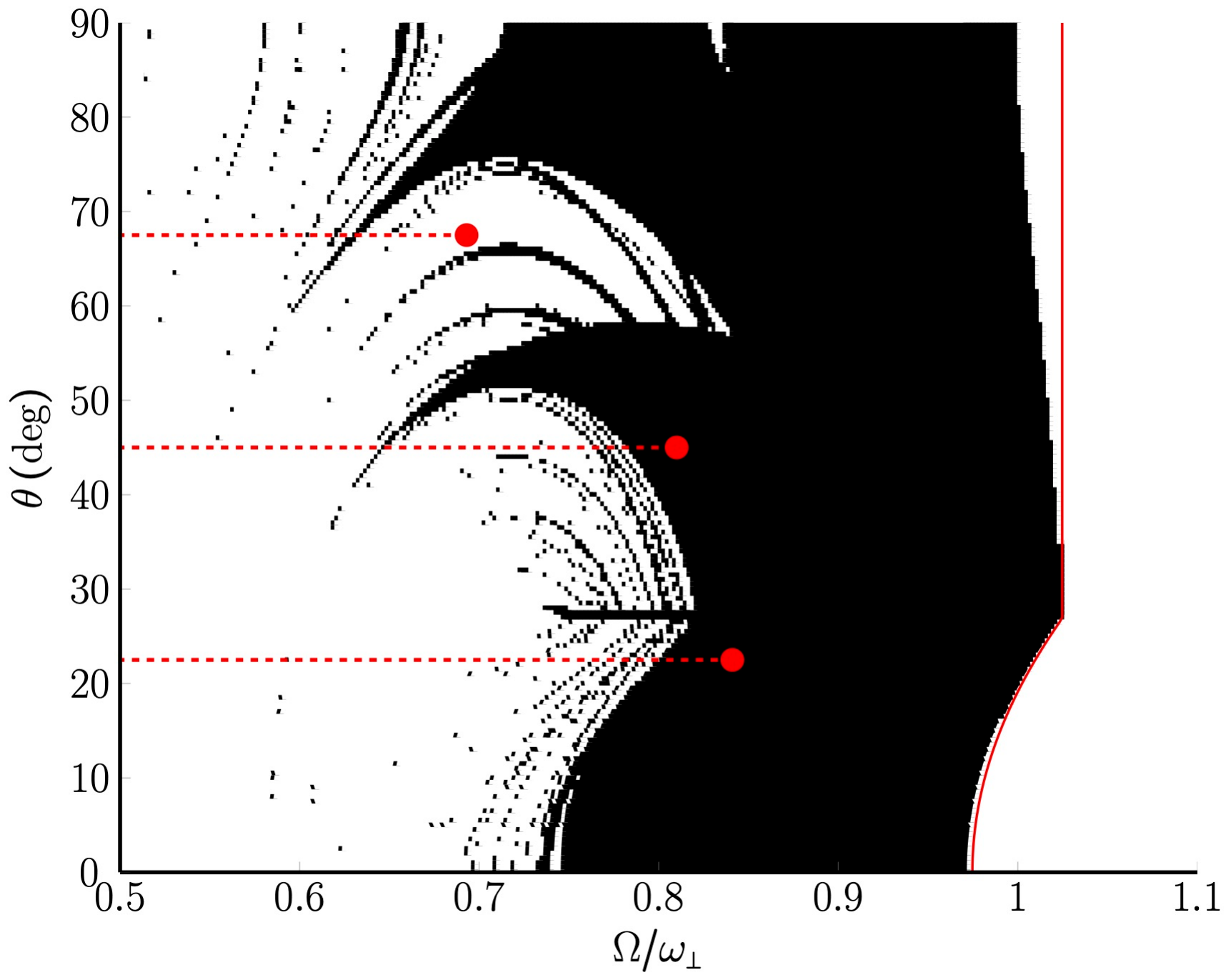}
\vspace*{-5mm}
\caption{Phase diagram of the dynamical stability of Branch I for $\varepsilon = 0.05,\,\gamma = 4/3$, with $N_{\text{max}} = 10$; Branch I is dynamically unstable at the shaded points of parameter space. The red dashed lines and markers denote the trajectories of the GPE simulations and the corresponding instability frequency, respectively. The red unbroken curve denotes the endpoints of Branch I, $\Omega = \min\lbrace\Omega_{xc},\Omega_{yc}\rbrace$.}
\label{epsilon0point05gamma4over3upperheaviside}
\end{figure}

We now proceed to describe the eigenvalues of the collective modes for Branch I of the stationary solutions presented in Sec.~\ref{sec:level3}. From an inspection of $\mathcal{M}$, every possible collective mode features the same maximum polynomial order for both $\delta n(\tilde{\mathbf{r}})$ and $\delta S(\tilde{\mathbf{r}})$, except for a spatially uniform phase fluctuation without a corresponding density fluctuation that is associated with a null eigenvalue. This is a manifestation of the Goldstone mode and is a consequence of the broken $\mathbb{U}(1)$ symmetry that characterizes Bose-Einstein condensation~\cite{pra_54_5_4204-4212_1996, pitaevskiistringaribec}. Fixing $N_{\text{max}} = 10$, we diagonalize Eq.~\eqref{eq:pertmatrix} over the discretely binned parameter space specified by the domain of Branch I of the stationary solutions described in Sec.~\ref{sec:level3.1} ($\varepsilon = 0$, $\gamma = 3/4$ and $\theta\in [0^{\circ}, 90^{\circ}]$). In Fig.~\ref{epsilon0gamma0point75upperheaviside}, we have shaded the bins where the respective Branch I solutions are associated with at least one eigenvalue of $\mathcal{M}$ with a real positive component. To linear order, these points in parameter space comprise a domain of guaranteed dynamical instability. A similar diagonalization of $\mathcal{M}$ with respect to the stationary solutions in Sec.~\ref{sec:level3.2}, i.e. $\varepsilon = 0.05$, $\gamma = 4/3$ and $\theta\in [0^{\circ}, 90^{\circ}]$, yields the stability diagram depicted in Fig.~\ref{epsilon0point05gamma4over3upperheaviside}.


From Figs.~\ref{epsilon0gamma0point75upperheaviside} and \ref{epsilon0point05gamma4over3upperheaviside}, we can see that Branch I is stable for small rotation frequencies and becomes dynamically unstable as $\Omega\rightarrow\omega_{\perp}$. In both cases, we find that the first rotation frequency of instability is lower for larger trap tilt angles, which we attribute to the effective ellipticity of the trapping in the upright co-rotating frame becoming larger as $\theta \rightarrow \pi/2$. Due to the high order of polynomial perturbations that is required to realize unstable collective modes in the limit $\Omega\rightarrow\min\lbrace\Omega_{xc},\Omega_{yc}\rbrace$, Figs.~\ref{epsilon0gamma0point75upperheaviside} and \ref{epsilon0point05gamma4over3upperheaviside} erroneously predict a region of dynamical stability. This limit is represented in Figs.~\ref{epsilon0gamma0point75upperheaviside} and \ref{epsilon0point05gamma4over3upperheaviside} by the red lines that plot $\min\lbrace\Omega_{xc},\Omega_{yc}\rbrace$ as a function of $\theta$; for a sufficiently large value of $N_{\text{max}}$, these red lines would be the boundary of the domain of dynamical instability.

\section{\label{sec:level5}Gross-Pitaevskii Equation Simulations}
In the preceding two sections we have found that the rotation of a tilted harmonic trap induces a nontrivial tilted angle of the condensate's density profile and that the stationary solutions become dynamically unstable as $\Omega\rightarrow\min\lbrace\Omega_{xc}, \Omega_{yc}\rbrace$. However, the Thomas-Fermi approximation does not provide information about the behavior of the condensate after the dynamical instability has manifested itself, nor does it predict whether or not the narrow regions of instability that extend to lower rotation frequencies in Figs.~\ref{epsilon0gamma0point75upperheaviside} and \ref{epsilon0point05gamma4over3upperheaviside} are negligible during a quasi-adiabatic rampup of $\Omega$. In order to attempt to answer these questions, we have also directly explored this system via numerically solving the GPE and thereby simulating a quasi-adiabatic rampup of a harmonic trap's rotation frequency from zero. In this section, we employ the same set of trapping parameters that were specified in the discussion of the TF stationary states and their dynamical stability, i.e. $\theta \in \lbrace\pi/8,\pi/4,3\pi/8\rbrace$ with either $\lbrace\gamma = 3/4,\,\varepsilon = 0\rbrace$ or $\lbrace\gamma = 4/3,\,\varepsilon = 0.05\rbrace$, and discuss the results of the GPE simulations.

Our procedure for solving Eq.~\eqref{eq:rescaledgpe} in the upright, co-rotating coordinate frame (denoted by $\mathbf{r}$) is as follows. We set the rescaled two-body interaction strength as $\tilde{g} = 10^4$ and specify a $200\times 200\times 200$ spatial grid with the intervals $\Delta x = \Delta y = \Delta z = 0.25l_{\perp}$; these parameters are sufficient for the ground state at $\Omega = 0$ to be well-described by the TF stationary solution. Initially, the backward Euler method is utilized to simulate Eq.~\eqref{eq:rescaledgpe} in imaginary time, with a suitable trial state as the initial condition, and the converged solution is taken as the ground state solution at zero rotation~\cite{kinetrelatmod_6_1_1-135_2013}. Before propagating this resulting solution in real time, the local value of the condensate density is randomly perturbed by up to $5\%$ of the original value in order to represent the environmental noise or experimental imperfections that would seed any potentially unstable collective modes. This perturbed state is used as the initial condition for the real-time evolution of the GPE, which is achieved using the Alternate Direction Implicit-Time Splitting pseudoSPectral (ADI-TSSP) Strang scheme~\cite{bao_wang_2006}. We employ a timestep of $\Delta t = 0.004\omega_{\perp}^{-1}$ and an angular acceleration $\frac{\Delta\Omega}{\Delta t} = 0.0005\omega_{\perp}^2$, where the resulting increase in $\Omega$ at each timestep, $\Delta\Omega = 2\times 10^{-6}$, is sufficiently small that the condition of adiabaticity holds. Therefore, the condensate is expected to smoothly follow Branch I of the TF stationary solutions during the rampup procedure. During the real-time evolution of the GPE, we extract the observables $R_x$, $R_y$, $R_z$, and $\xi$ by fitting the density at $x = z = 0$ to the $1$D TF density profile, $n(y) = n_0(1 - y^2/R_y^2)$, and similarly the density at $y = 0$ to the $2$D TF density profile, $n(x, z) = n_0\lbrace 1 - x^2[\cos^2(\theta - \xi)/R_x^2 + \sin^2(\theta-\xi)/R_z^2] - z^2[\sin^2(\theta - \xi)/R_x^2 + \cos^2(\theta-\xi)/R_z^2] - (1/R_x^2-1/R_z^2)\sin[2(\theta-\xi)]xz\rbrace$. Note that the form of these density cross-sections can be found by applying the transformation in Eq.~\eqref{eq:adjusttiltframe} to Eq.~\eqref{eq:nstat}. Subsequently, we may determine $\alpha$ and $\delta$ via Eqs.~\eqref{eq:alphadefn} and \eqref{eq:deltadefn}.

\subsection{\label{sec:level5.1}Prolate, Symmetric Trapping}
\begin{figure}
  \includegraphics[width=\linewidth]{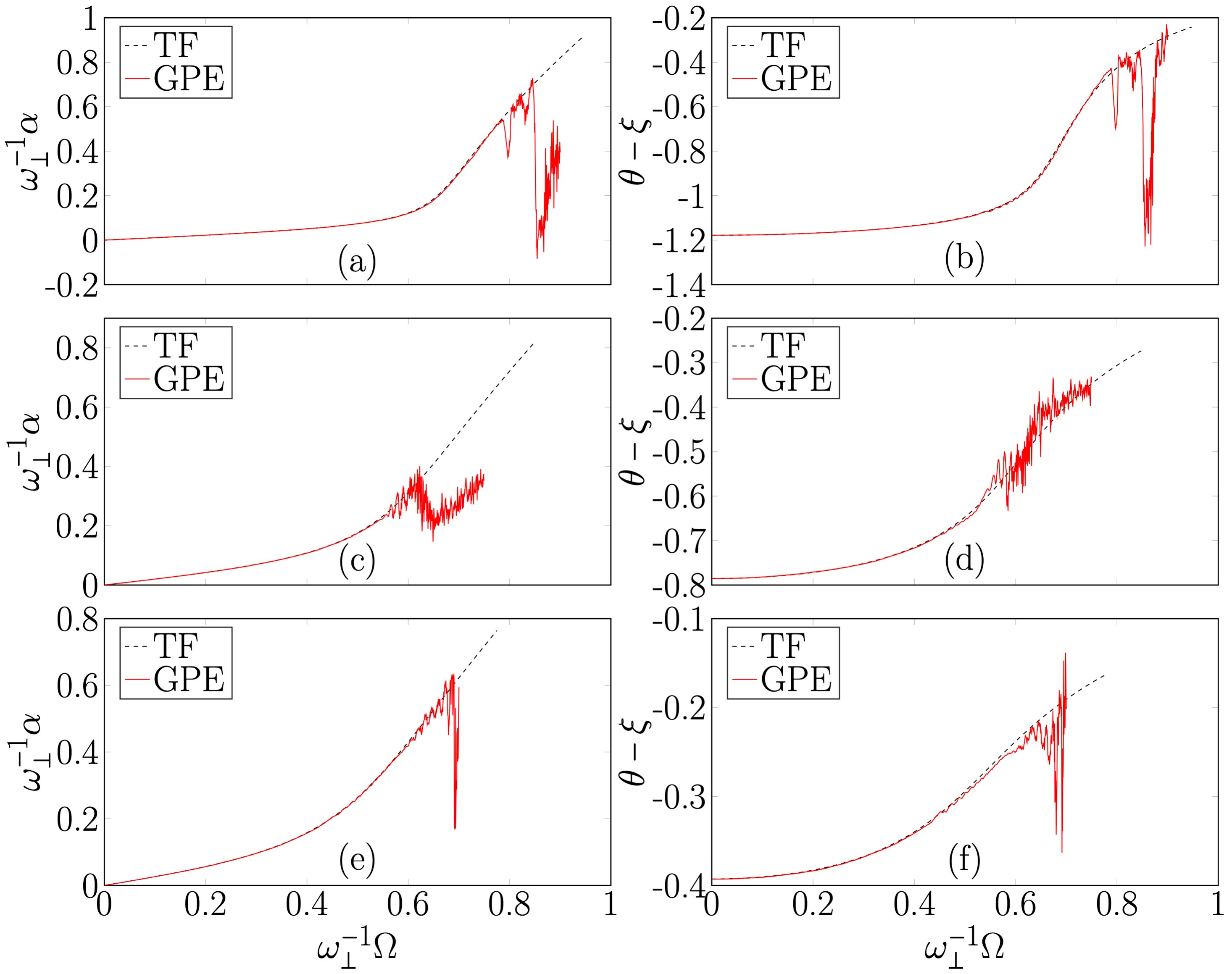}
  \vspace*{-5mm}
  \caption{Comparison of the TF stationary solutions along Branch I to GPE simulations, for $\varepsilon = 0$ and $\gamma = 3/4$, during a quasi-adiabatic rampup of $\Omega$. The first column (a, c, e) plots $\alpha$ as a function of $\Omega$ and the column (b, d, f) plots $\theta - \xi$ as a function of $\Omega$, where $\theta$ is equal to $\pi/8$ (a, b), $\pi/4$ (c, d), or $3\pi/8$ (e, f).}
  \label{case_1_alphatmx}
\end{figure}

In Fig.~\ref{case_1_alphatmx} we compare $\alpha$ and $\xi$ as obtained from the GPE simulations of a quasi-adiabatic rampup of $\Omega$, when $\varepsilon = 0$ and $\gamma = 3/4$, to the TF results in Fig.~\ref{epsilon0gamma0point75thetaalltf}(a) and (b). Here, the first (a, c, e) and second (b, d, f) columns correspond to $\alpha$ and $\theta - \xi$, respectively, as functions of $\Omega$ while the rows correspond to distinct tilting angles: $\theta = \pi/8$ in the first row (a, b), $\theta = \pi/4$ in the second row (c, d), and $\theta = 3\pi/8$ in the third row (e, f). Figure \ref{case_1_alphatmx} demonstrates that the condensate initially follows the TF stationary state closely during the quasi-adiabatic acceleration of the rotation frequency, which confirms the prediction in Fig.~\ref{epsilon0gamma0point75upperheaviside} that the TF stationary states are dynamically stable for low rotation frequencies. However, as $\Omega\rightarrow\omega_{\perp}$, each of the trajectories from the numerical simulations diverge dramatically from the TF-based predictions. This indicates the onset of a dynamical instability, as predicted in Fig.~\ref{epsilon0gamma0point75upperheaviside}, where the condensate has been forced away from the TF stationary state due to the uncontrolled growth of collective modes. Similar behavior is seen in the analogous comparison of the TF- and GPE-derived values of $\delta$, which we have included in Appendix~\ref{sec:level9} for the reader's reference.

The rotation frequencies at which the condensate densities in each of the three simulations diverge from the corresponding TF stationary state densities are depicted as red circular markers in Fig.~\ref{epsilon0gamma0point75upperheaviside}. When $\theta = \pi/8$ or $3\pi/8$, the onset of dynamical instability agrees well with the predictions of the linearized hydrodynamical formalism. However, when $\theta = \pi/4$ the critical rotation frequency is approximately $0.55\omega_{\perp}$, whereas Fig.~\ref{epsilon0gamma0point75upperheaviside} predicts that the stationary solution is always unstable when $\Omega \gtrsim 0.7\omega_{\perp}$. We attribute this discrepancy to the existence of the small fringes of dynamical instability that intersect the trajectory of the $\theta = \pi/4$ simulation at $\Omega \approx 0.50\omega_{\perp}$ and $0.51\omega_{\perp}$, which are sufficient to destabilize the stationary solution. The fringes at $\Omega \approx 0.57\omega_{\perp}$ and $\Omega \approx 0.44\omega_{\perp}$ that are crossed by the trajectories of the simulations for $\theta = \pi/8$ and $\theta = 3\pi/8$, respectively, seem to be too narrow to sufficiently destabilize the stationary states. While another such fringe is crossed by the simulation for $\theta = \pi/8$ when $\Omega \approx 0.51\omega_{\perp}$, and a set of fringes is crossed by the $\theta = 3\pi/8$ simulation when $\Omega \approx 0.72\omega_{\perp}$, they are very close to the continuous domain of dynamical instability and thus their effect is relatively minimal. Furthermore, the GPE simulations also capture nonlinear effects that are ignored in the the linearized hydrodynamic formalism.

\begin{figure}
  \includegraphics[width=\linewidth]{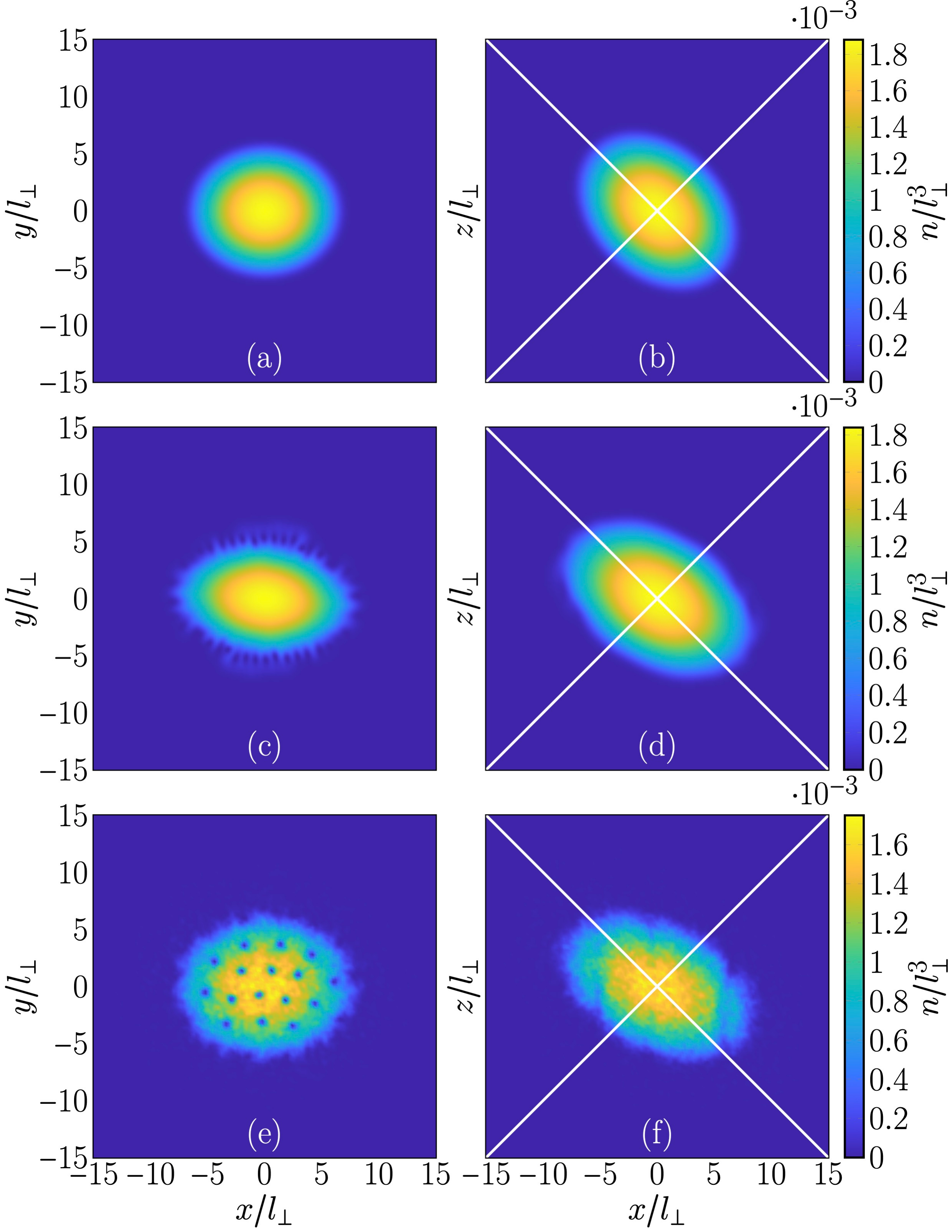}
  \vspace*{-5mm}
  \caption{Cross-sections of the condensate density in the co-rotating $x$-$y$ (first column) and $x$-$z$ (second column) planes for $\varepsilon = 0$, $\gamma = 3/4$, and $\theta = \pi/4$, during a quasi-adiabatic rampup of $\Omega$ at $\Omega = 0.25\omega_{\perp}$ (first row) and $\Omega = 0.575\omega_{\perp}$ (second row), and after $500\omega_{\perp}^{-1}$ at constant $\Omega = 0.575\omega_{\perp}$ (third row). The white lines represent the co-rotating $X$-$Z$ axes.}
  \label{case1_45}
\end{figure}

The deviation of the simulations from the respective TF stationary states at higher rotation frequencies can be better understood by examining the cross-sections of the condensate densities at $y = 0$ and $z = 0$, which provide us information about the density profiles; tilting angles, $\xi$, and the semi-axes, $\lbrace R_x,\,R_y,\,R_z\rbrace$, during the acceleration of the rotation frequency. These cross-sections are presented in Fig.~\ref{case1_45} for the trap tilt $\theta = \pi/4$ when $\Omega$ equals $0.25\omega_{\perp}$ (first row) and $0.575\omega_{\perp}$ (second row). When $\Omega = 0.575\omega_{\perp}$, we halt the rampup of $\Omega$ and then evolve the GPE at constant rotation frequency for a duration of $500\omega_{\perp}^{-1}$; the cross-sections at the end of this procedure are given in the third row of Fig.~\ref{case1_45}. We include the results for the analogous procedures performed with the same trapping geometry but with $\theta = \lbrace\pi/8,3\pi/8\rbrace$ in Appendix~\ref{sec:level9}. In order to aid the reader's visualization of how the density's principal axes do not generally coincide with those of either the trapping frame or the rotation axis, the $X$-$Z$ Cartesian axes, i.e. the principal axes of the trapping, are overlaid in white upon the cross-sections at $y = 0$.

In Fig.~\ref{case1_45} we can see that the density profile is smooth when the condensate is dynamically stable against the initially seeded perturbation and, as predicted by the TF theory, its symmetry axes in the $x-z$ plane are slightly tilted away from those of the trap. However, when the condensate initially enters the regime of dynamic instability, the density develops surface ripples and a surrounding cloud as some of the atoms are ejected from the centre of the condensate, as seen in the second column of Fig.~\ref{case1_45}. Moreover, we see that after evolution over a period of $500\omega_{\perp}^{-1}$ at constant rotation frequency, $\Omega = 0.575\omega_{\perp}$, the condensate does not resemble a smooth TF distribution but has been subject to quantum vortex nucleation after further atoms have been ejected from the centre of the condensate. This behaviour is a well-known phenomenon that occurs in the rotation of an upright, anisotropic harmonic trap containing a BEC~\cite{prl_87_19_190402_2001, prl_86_20_4443-4446_2001, pra_65_2_023603_2002, pra_67_3_033610_2003, prl_92_2_020403_2004, prl_95_14_145301_2005, pra_73_6_061603r_2006, jphysb_40_18_3615-3628_2007} and thus it is not surprising that it occurs in this system. Crucially, an inspection of Fig.~\ref{case1_45}(f) shows that the vortex lines coincident upon the $x-z$ plane are almost completely aligned along the rotation axis. This is in contrast to the background condensate density whose symmetry axes are tilted with respect to both $\lbrace\hat{x}, \hat{z}\rbrace$ and $\lbrace\hat{X}, \hat{Z}\rbrace$. While the vortices that are seen in Fig.~\ref{case1_45} are not ordered in a lattice, we expect that after a considerably longer period of evolution of the GPE at a constant rotation frequency, the final state of the system is a triangular Abrikosov vortex lattice, as is seen in BECs subject to rotation about a principal axis of the trapping~\cite{science_292_5516_476-479_2001, prl_92_2_020403_2004, prl_95_14_145301_2005, pra_73_6_061603r_2006}.

\subsection{\label{sec:level5.2}Oblate, Asymmetric Trapping}
\begin{figure}
  \includegraphics[width=\linewidth]{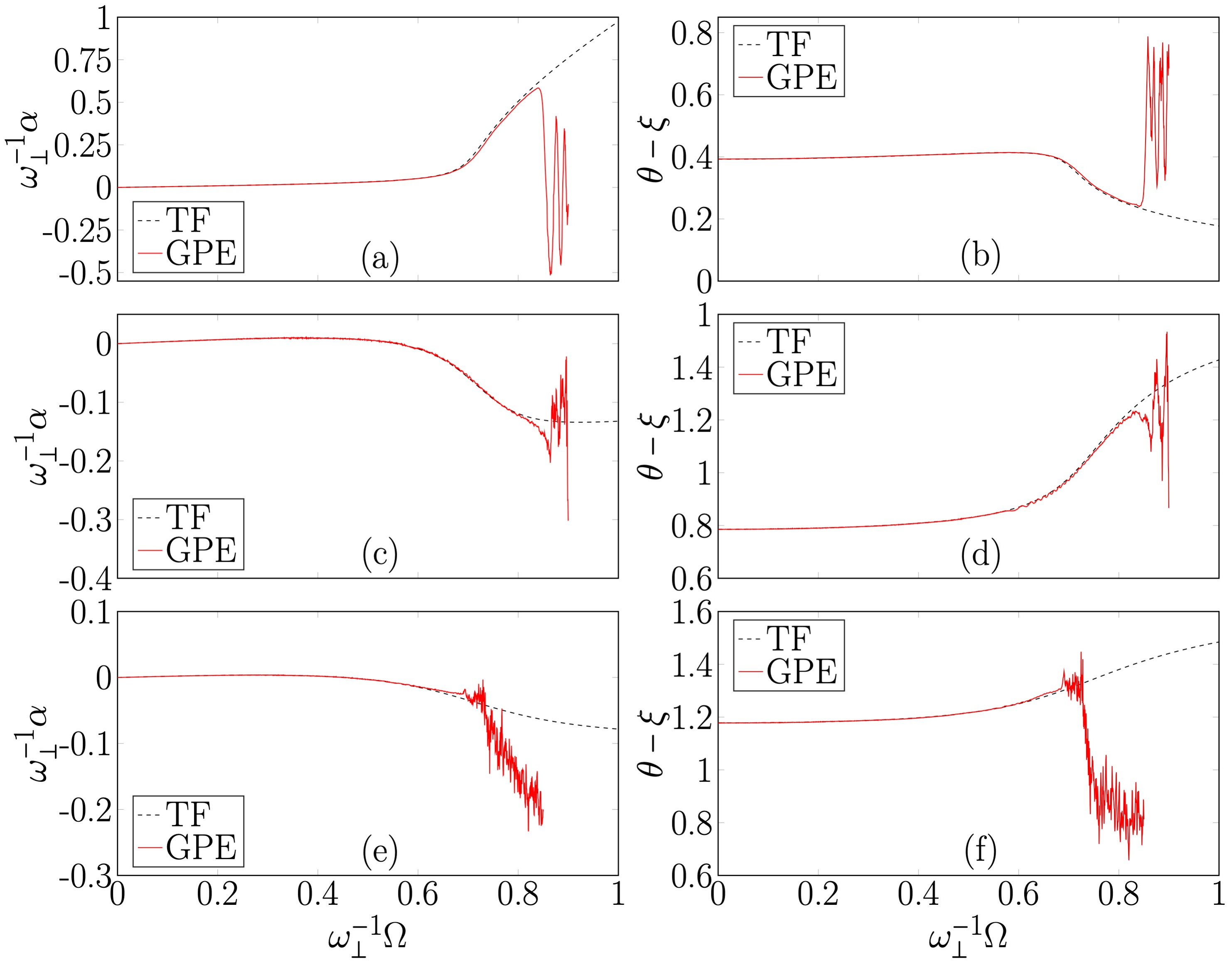}
  \vspace*{-5mm}
  \caption{Comparison of the TF stationary solutions along Branch I to GPE simulations, for $\varepsilon = 0.05$ and $\gamma = 4/3$, during a quasi-adiabatic rampup of $\Omega$. The first column (a, c, e) plots $\alpha$ as a function of $\Omega$ and the column (b, d, f) plots $\theta - \xi$ as a function of $\Omega$, where $\theta$ is equal to $\pi/8$ (a, b), $\pi/4$ (c, d), or $3\pi/8$ (e, f).}
  \label{case_2_alphatmx}
\end{figure}

We now describe the results of the analogous GPE simulations for a trap with the parameters $\gamma = 4/3$ and $\varepsilon = 0.05$. In Fig.~\ref{case_2_alphatmx} we compare $\alpha$ and $\theta - \xi$ from these GPE simulations to the TF results in Fig.~\ref{epsilon0point05gamma4over3thetaalltf}(a) and (b). Here, the first (a, c, e) and second (b, d, f) columns correspond to $\alpha$ and $\theta - \xi$, respectively, as functions of $\Omega$ while the rows correspond to distinct tilting angles: $\theta = \pi/8$ in the first row (a, b), $\theta = \pi/4$ in the second row (c, d), and $\theta = 3\pi/8$ in the third row (e, f). Just as in the simulations described in Sec.~\ref{sec:level5.1}, the condensate is seen to be unstable at higher rotation frequencies against collective modes seeded by the random perturbation at $t = 0$. This agrees with the behavior seen in a comparison of the semi-analytically and numerically obtained values of $\delta$, which we have included in Appendix~\ref{sec:level9}. A comparison may also be made with the prediction of dynamical instability in Fig.~\ref{epsilon0point05gamma4over3upperheaviside}, where we have indicated the rotation frequencies at which the GPE states diverge considerably from the TF states via red circular markers. When $\theta \in \lbrace\pi/8, \pi/4\rbrace$, these rotation frequencies are greater than the respective threshold frequencies above which the stationary states are always dynamically unstable. However, for $\theta = 3\pi/8$, the rotation frequency where the GPE solution diverges wildly from the TF prediction occurs at $\Omega \approx 0.69\omega_{\perp}$, which is considerably lower than the prediction of Fig.~\ref{epsilon0point05gamma4over3upperheaviside} that the stationary state is dynamically unstable when $\Omega \gtrsim 0.80\omega_{\perp}$. This may be attributed to the fact that the trajectory of the quasi-adiabatic rampup crosses a fringe of dynamical instability when $\Omega \approx 0.63\omega_{\perp}$. We note that a similar fringe is crossed when $\theta = \pi/4$ and $\Omega \approx 0.65\omega_{\perp}$, but this fringe is narrower than the one that destabilizes the $\theta = 3\pi/8$ stationary state. While the quasi-adiabatic trajectory for $\theta = \pi/8$ crosses several narrow fringes when $\Omega\in(0.75\omega_{\perp},0.78\omega_{\perp})$, their effect is relatively minimal as they are closely followed by the threshold for dynamical instability at $\Omega \approx 0.8\omega_{\perp}$.

\begin{figure}
  \includegraphics[width=\linewidth]{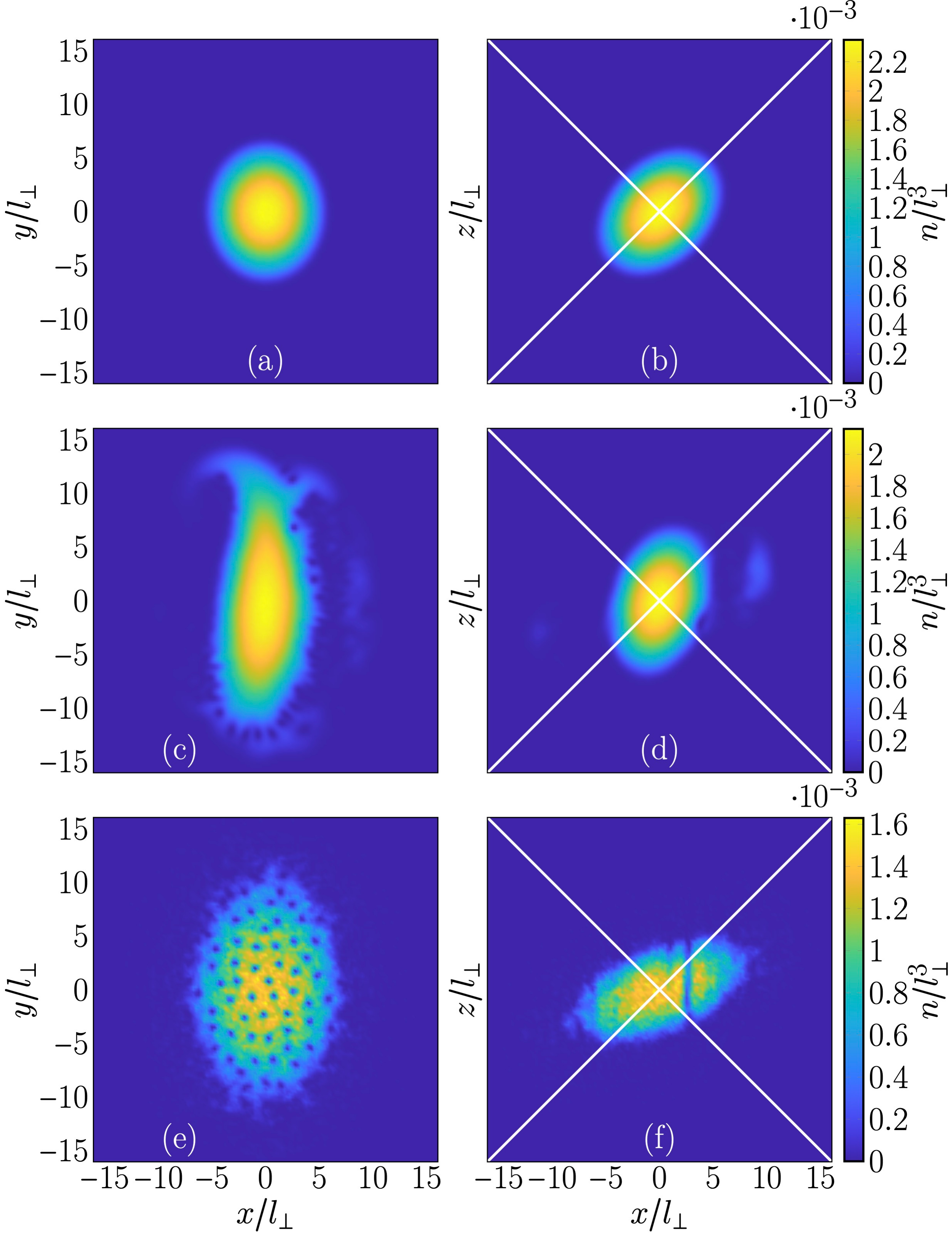}
  \vspace*{-5mm}
  \caption{Cross-sections of the condensate density in the co-rotating $x$-$y$ (first column) and $x$-$z$ (second column) planes for $\varepsilon = 0.05$, $\gamma = 4/3$, and $\theta = \pi/4$, during a quasi-adiabatic rampup of $\Omega$ at $\Omega = 0.4\omega_{\perp}$ (first row) and $\Omega = 0.85\omega_{\perp}$ (second row), and after $500\omega_{\perp}^{-1}$ at constant $\Omega = 0.85\omega_{\perp}$ (third row). The white lines represent the co-rotating $X$-$Z$ axes.}
  \label{case2_45}
\end{figure}

We can also visualize the GPE solutions for $\theta = \pi/4$ by plotting the cross-sections of the density in the $x$-$y$ and $x$-$z$ planes for $\theta = \pi/4$ in Fig.~\ref{case2_45}, with the corresponding plots for $\theta = \lbrace\pi/8, 3\pi/8\rbrace$ included for reference in Appendix~\ref{sec:level9}. Just as in Sec.~\ref{sec:level5.1}, the density cross-sections are smooth and ellipsoidal at low rotation frequencies during a quasi-adiabatic rampup of $\Omega$. This is evident in the first row of Fig.~\ref{case2_45}, where $\Omega = 0.4\omega_{\perp}$, which also shows that the condensate density's principal axes are slightly tilted away from those of the trapping. Similarly, we again observe that the onset of the dynamical instability is marked by the presence of a high-density core with surface rippling, surrounded by a low-density halo-like cloud, in the second row of Fig.~\ref{case2_45} where $\Omega = 0.85\omega_{\perp}$. Upon halting the acceleration of the rotation frequency when $\Omega = 0.85\omega_{\perp}$ and then evolving the GPE at constant rotation frequency for the duration $500\omega_{\perp}^{-1}$, the condensate is subject to the nucleation of a large number of vortices as seen in the third row of Fig.~\ref{case2_45}. More vortices are found in Fig.~\ref{case2_45} than in Fig.~\ref{case1_45}, which is likely due to the higher rotation frequency at which the quasi-adiabatic rampup was halted. In both cases, however, we find that the vortex lines coincident upon the $x-z$ plane are almost completely aligned along the $z$-axis and that the background condensate density profile is tilted with respect to both the rotating trap and the rotation axis.

\section{\label{sec:level6}Conclusion}
In this work, we have extended the Thomas-Fermi theory for slowly rotating Bose-Einstein condensates in anisotropic harmonic traps to account for rotations of the trap about an axis that is not one of its three principal axes. In traps subject to tilted rotation, the stationary state density profile's principal axes are generally tilted with respect to those of both the confinement and the rotation. The quadrupolar irrotational velocity profile describing the vorticity-free flow of the condensate is also modified as a consequence of the tilting of the rotating harmonic trap. Our analysis of the resulting stationary solutions demonstrate the existence of previously unknown, tilted, solution branches (Branches III and IV) that exist when $\Omega > \omega_{\perp}$. Although we have only conducted a systematic study of the dynamical stability of one of the five stationary solution branches, Branch I, it is nonetheless interesting to consider whether Branch III, in particular, becomes dynamically unstable immediately upon reaching $\Omega = \omega_{\perp}$ or if its stability persists for a larger window. When $\theta = 0$ and $\varepsilon \neq 0$, a method that has been proposed for accessing the branch defined for $\Omega \in [\omega_{\text{b1}},\infty)$ is to start from the $\alpha = 0$ stationary solution when $\varepsilon = 0$ and then quasi-adiabatically tune $\varepsilon$ to the desired final value whilst keeping $\Omega$ fixed~\cite{prl_86_3_377-380_2001}. In principle a similar method could be utilized to explore the stationary solution along Branch III for $\Omega\in\left[\max\lbrace\Omega_{xc},\Omega_{yc}\rbrace, \Omega_{\text{b}2}\right]$ and $\theta \neq 0$, in an anisotropic harmonic trap, by starting from an isotropic trap rotating at a fixed frequency and adiabatically tuning its anisotropy as desired.

Our work also suggests that vortices are nucleated in response to a tilted rotating trap and are aligned along the rotation axis, and not along one of the tilted principal axes of the trap. Although we expect that the condensate's final state in the dynamically unstable domain to be a triangular vortex lattice, further work in this direction is needed to resolve this, as well as the tilting angle of the background condensate density and the response of the vortices to perturbations~\cite{pra_62_6_063617_2000, prl_86_21_4725-7428_2001, prl_90_10_100403_2003, prl_91_9_090403_2003, prl_91_10_100402_2003, prl_91_11_110402_2003, prl_93_8_080406_2004, prl_101_2_020402_2008, prl_113_16_165303_2014}. The formalism outlined here for finding rotating frame stationary solutions with a tilting of the trap's symmetry axes can be extended to more exotic condensates than the scalar one we have considered. Notably, in the field of dipolar quantum gases, we expect that dipolar Bose-Einstein condensates in the TF limit can be described in a similar manner, based on previous work on rotating either the trapping or the dipole polarization about a principal axis of the trapping~\cite{prl_98_15_150401_2007, pra_80_3_033617_2009, jphyscondesmatter_29_10_103004_2017, prl_122_5_050401_2019, pra_100_2_023625_2019}. Similarly we would expect that spin-orbit-coupled BECs subject to an artificial gauge field that induces a synthetic rotation about a non-principal axis would be described analogously, in the TF limit, to the formalism we have introduced here~\cite{pra_84_2_021604r_2011, prl_120_18_183202_2018}.

\begin{acknowledgments}
S. B. P. is supported by an Australian Government Research Training Program Scholarship and by the University of Melbourne. The numerical simulations were conducted on the Spartan HPC cluster~\cite{meade2017spartan}, and we thank Research Computing Services at the University of Melbourne for access to this resource. We also thank Nick Parker and Thomas Bland for several stimulating discussions that motivated this work.
\end{acknowledgments}

\appendix
\section{\label{sec:level8}Endpoints of the Stationary Solution Branches}
The endpoints of the branches, save for Branch V, are defined by the limits given by:
\begin{enumerate}[label=\alph*)]
\item \begin{center}$\tilde{\omega}_x \rightarrow 0$ and $\tilde{\omega}_y,\text{}\tilde{\omega}_z \neq 0$,\end{center} \label{en:endpointxapp}
\item \begin{center}$\tilde{\omega}_y \rightarrow 0$ and $\tilde{\omega}_x,\text{}\tilde{\omega}_z \neq 0$,\end{center} \label{en:endpointyapp}
\item \begin{center}$\tilde{\omega}_x,\text{}\tilde{\omega}_y \rightarrow 0$ and $\tilde{\omega}_z \neq 0$,\end{center} \label{en:endpointxyapp}
\item \begin{center}$\tilde{\omega}_y,\text{}\tilde{\omega}_z \rightarrow 0$ and $\tilde{\omega}_x \neq 0$.\end{center} \label{en:endpointyzapp}
\end{enumerate}
Let us denote the critical values of quantities such as $\Omega$, $\alpha$, $\delta$, and $\xi$ at the limiting cases~\ref{en:endpointxapp}, \ref{en:endpointyapp}, \ref{en:endpointxyapp} and \ref{en:endpointyzapp} by the subscripts $xc$, $yc$, $xyc$, and $yzc$ respectively.

For case~\ref{en:endpointxapp}, Eq.~\eqref{eq:alphaeqn} implies that we have $\alpha_{xc} = \Omega_{xc}\cos(\theta - \xi_{xc})$. Substituting this into Eqs.~\eqref{eq:omegaxeff} and \eqref{eq:xzcoeffzero} yields
\begin{align}
  \Omega_{xc}^2\cos^2(\theta-\xi_{xc}) &= (1-\varepsilon)\cos^2\xi_{xc} + \gamma^2\sin^2\xi_{xc}, \label{eq:omxzerolim1} \\
  \Omega_{xc}^2\sin[2(\theta-\xi_{xc})] &= (\gamma^2 -1+\varepsilon)\sin(2\xi_{xc}). \label{eq:omxzerolim2}
\end{align}
Equations \eqref{eq:omxzerolim1} and \eqref{eq:omxzerolim2} admit the solution pair
\begin{align}
  \Omega_{xc}^2 &= \frac{\gamma^2(1-\varepsilon)\omega_{\perp}^2}{\gamma^2\cos^2\theta + (1-\varepsilon)\sin^2\theta}, \label{eq:omxzerolimOm} \\
  \cos^2\xi_{xc} &= \frac{\gamma^4\cos^2\theta}{\gamma^4\cos^2\theta + (1-\varepsilon)^2\sin^2\theta}, \label{eq:omxzerolimxi}
\end{align}
which together yield the solution for $\alpha_{xc}$ via $\alpha_{xc} = \Omega\cos(\theta - \xi_{xc})$. We may also solve for $\delta_{xc}$ by substituting these roots into Eq.~\eqref{eq:deltaeqn}. Crucially, when $\theta = 0$, Eqs.~\eqref{eq:omxzerolim1} and \eqref{eq:omxzerolim2} imply that $\xi_{xc} = \delta_{xc} = 0$ and $\Omega_{xc} = \alpha_{xc} = \omega_{\perp}\sqrt{1 - \varepsilon}$, as expected~\cite{prl_86_3_377-380_2001}. The same limiting forms are also valid when the trapping is axially symmetric about the $y$-axis, i.e. $\gamma = \sqrt{1-\varepsilon}$. For case~\ref{en:endpointyapp}, we find that we have
\begin{align}
  \alpha_{yc} &= -\Omega_{yc}\cos(\theta - \xi_{yc}), \label{eq:omyzeroaldel1} \\
  \delta_{yc} &= \Omega_{yc}\sin(\theta - \xi_{yc}). \label{eq:omyzeroaldel2}
\end{align}
Substitution of these into Eq.~\eqref{eq:omegayeff} yields
\begin{equation}
  \Omega_{yc} = \omega_{\perp}\sqrt{1 + \varepsilon}, \label{eq:omyzerolimOm}
\end{equation}
and thus
\begin{equation}
  3(1+\varepsilon)\sin[2(\theta-\xi_{yc})] = (1-\varepsilon-\gamma^2)\sin(2\xi_{yc}), \label{eq:omyzerolimxieqn}
\end{equation}
which admits the solution
\begin{equation}
  \tan(2\xi_{yc}) = \frac{3(1+\varepsilon)\sin(2\theta)}{3(1+\varepsilon)\cos(2\theta)-\gamma^2+1-\varepsilon}. \label{eq:omyzerolimxi}
\end{equation}
Via Eqs.~\eqref{eq:omyzeroaldel1} -- \eqref{eq:omyzerolimOm}, and \eqref{eq:omyzerolimxi}, we may obtain the solutions of $\alpha_{yc}$ and $\delta_{yc}$ in this limit. For the special cases that the trap is not tilted, i.e. $\theta = 0$, and/or is axially symmetric about the $y$-axis, i.e. $\gamma = \sqrt{1-\varepsilon}$, we have $\alpha_{yc} = -\Omega_{yc} = -\omega_{\perp}\sqrt{1+\varepsilon}$ and $\delta_{yc} = \xi_{yc} = 0$~\cite{prl_86_3_377-380_2001}.

The limits~\ref{en:endpointxyapp} and \ref{en:endpointyzapp} are somewhat more involved. In case~\ref{en:endpointxyapp} we have $\delta_{xyc} = \Omega_{xyc}\sin(\theta-\xi_{xyc})$, but the limit of $\alpha_{xyc}$ is not as obvious and must be found by solving Eq.~\eqref{eq:xzcoeffzero}. This gives us
\begin{equation}
  \alpha_{xyc} = \frac{\Omega_{xyc}^2\sin[2(\theta-\xi_{xyc})] + (\gamma^2 - 1 + \varepsilon)\sin(2\xi_{xyc})}{4\Omega_{xyc}\sin(\theta-\xi_{xyc})}. \label{eq:omxyzeroal}
\end{equation}
Substituting these relations into Eqs.~\eqref{eq:omegaxeff} and \eqref{eq:omegayeff} results in the system of equations given by:
\begin{widetext}
\begin{gather}
12\Omega^4\cos^2(\theta-\xi) - 8(\gamma^2+1-\varepsilon)\Omega^2 + (\gamma^2-1+\varepsilon)\left\lbrace 8\Omega^2\cos(2\xi) + \sin(2\xi)\left[\frac{4\Omega^2}{\tan(\theta-\xi)} - \frac{(\gamma^2-1+\varepsilon)\sin(2\xi)}{\sin^2(\theta-\xi)}\right]\right\rbrace = 0, \label{eq:omxyzerolimomxi1} \\
2\Omega^2\lbrace 8(1+\varepsilon)+\Omega^2[1+9\cos(2(\theta-\xi))]\rbrace\sin(\theta-\xi) + (\gamma^2-1+\varepsilon)\sin(2\xi)\left[12\Omega^2\cos(\theta-\xi) + \frac{(\gamma^2-1+\varepsilon)\sin(2\xi)}{\sin(\theta-\xi)}\right] = 0. \label{eq:omxyzerolimomxi2}
\end{gather}
\end{widetext}
Solving these simultaneously for $\Omega$ and $\xi$ yields the limiting values, $\Omega_{xyc}$ and $\xi_{xyc}$, which subsequently allows for the solution of $\alpha_{xyc}$ and $\delta_{xyc}$ from Eq.~\eqref{eq:omxyzeroal} and the relation $\delta_{xyc} = \Omega_{xyc}\sin(\theta-\xi_{xyc})$ respectively.

In case~\ref{en:endpointyzapp}, we have $\alpha = -\Omega\cos(\theta-\xi)$ and from solving Eq.~\eqref{eq:xzcoeffzero} we also find that
\begin{equation}
\delta_{yzc} = \frac{(1 - \varepsilon - \gamma^2)\sin(2\xi_{yzc}) - \Omega_{yzc}^2\sin[2(\theta-\xi_{yzc})]}{4\Omega_{yzc}\cos(\theta-\xi_{yzc})}. \label{eq:omyzzerodel}
\end{equation}
The substitution of these relations into Eqs.~\eqref{eq:omegayeff} and \eqref{eq:omegazeff} results in the following system of equations:
\begin{widetext}
\begin{gather}
2\Omega^2\lbrace 8(1+\varepsilon)+\Omega^2[1-9\cos(2(\theta-\xi))]\rbrace\cos(\theta-\xi) + (\gamma^2-1+\varepsilon)\sin(2\xi)\left[12\Omega^2\sin(\theta-\xi) + \frac{(\gamma^2-1+\varepsilon)\sin(2\xi)}{\cos(\theta-\xi)}\right] = 0, \label{eq:omyzzerolimomxi1} \\
\gamma^2\cos^2\xi + (1-\varepsilon)\sin^2\xi - \Omega^2\sin^2(\theta-\xi) - \frac{(\gamma^2-1+\varepsilon)\sin(2\xi)\tan(\theta-\xi)}{2} + \left[\frac{\Omega\sin(\theta-\xi)}{2}+\frac{(\gamma^2-1+\varepsilon)\sin(2\xi)}{4\Omega\cos(\theta-\xi)}\right]^2 = 0. \label{eq:omyzzerolimomxi2}
\end{gather}
\end{widetext}
As in case~\ref{en:endpointxyapp}, solving these equations for $\Omega$ and $\xi$ yields $\Omega_{yzc}$ and $\xi_{yzc}$, and thus also $\alpha_{yzc}$ and $\delta_{yzc}$ via the relation $\alpha_{yzc} = -\Omega_{yzc}\cos(\theta-\xi_{yzc})$ and Eq.~\eqref{eq:omyzzerodel} respectively.

For both limits~\ref{en:endpointxyapp} and \ref{en:endpointyzapp}, the limits for the special case where $\theta = 0$ evaluate to simple closed forms given by:
\begin{align}
  \Omega_{xyc} = \Omega_{yzc} &= \omega_{\perp}\sqrt{1+\gamma^2+\sqrt{4\gamma^2+\varepsilon^2}}, \label{eq:omxyyzzerolimomthetazero} \\
  \cos^2\xi_{xyc} &= \frac{(2+\varepsilon)(\sqrt{4\gamma^2+\varepsilon^2}-\varepsilon)}{2(\gamma^2-1+\varepsilon)[\gamma^2-2(2+\varepsilon)]} \nonumber \\
  &-\frac{\gamma^2(\sqrt{4\gamma^2+\varepsilon^2}+1+\varepsilon-\gamma^2)}{(\gamma^2-1+\varepsilon)[\gamma^2-2(2+\varepsilon)]}, \label{eq:omxyzerolimxithetazero} \\
  \cos^2\xi_{yzc} &= \frac{\gamma^2(\sqrt{4\gamma^2+\varepsilon^2}-4)}{(\gamma^2-1+\varepsilon)[\gamma^2-2(2+\varepsilon)]} \nonumber \\
  &-\frac{(2+\varepsilon)(\sqrt{4\gamma^2+\varepsilon^2}-4+3\varepsilon)}{2(\gamma^2-1+\varepsilon)[\gamma^2-2(2+\varepsilon)]}, \label{eq:omyzzerolimxithetazero}
\end{align}
From these, $\alpha$ and $\delta$ may be evaluated in closed form in the respective limits. It is noted that Eqs.~\eqref{eq:omxyzerolimxithetazero} and \eqref{eq:omyzzerolimxithetazero} formally exhibit a removable singularity when the trap is axially symmetric about $\hat{y}$, i.e. $\gamma^2 = 1 - \epsilon$, and in this limit we have $\cos^2\xi_{xyc} = 2(1-\epsilon)/[3(2-\epsilon)]$ and $\cos^2\xi_{yzc} = (4-\epsilon)/[3(2-\epsilon)]$.

\section{\label{sec:level7}Visualizing the TF Density Profiles}
In this section, we provide the reader with a description of how the signs of the velocity amplitudes, $\alpha$ and $\delta$, and the angle $\theta - \xi$ provide us with a considerable amount of qualitative information of the shape of the Thomas-Fermi density profile corresponding to a given solution of Eqs.~\eqref{eq:tfratiosols} --~\eqref{eq:deltaeqn}. Let us restate the definitions of $\alpha$ and $\delta$ in terms of the TF semi-axes:
\begin{align}
\alpha &= \left(\frac{R_x^2 - R_y^2}{R_x^2 + R_y^2}\right)\Omega\cos(\theta-\xi), \label{eq:alphaappendixdefn} \\
\delta &= \left(\frac{R_y^2 - R_z^2}{R_y^2 + R_z^2}\right)\Omega\sin(\theta-\xi). \label{eq:deltaappendixdefn}
\end{align}
Since Eq.~\eqref{eq:nstat} exhibits a twofold rotation symmetry about the $y$-axis, $\xi$ has a period of $\pi$ and so we assume that $\theta - \xi \in (-\pi/2,\pi/2]$ without loss of generality. This choice of the principal branch fixes $\cos(\theta - \xi) \geq 0$, whereas $\sin(\theta - \xi) > 0$ when $\xi < \theta$ and $\sin(\theta - \xi) < 0$ when $\xi > \theta$. Therefore, by inspection of Eqs.~\eqref{eq:alphaappendixdefn} and \eqref{eq:deltaappendixdefn}, we have:
\begin{enumerate}[label=\roman*)]
\item $R_x>R_y>R_z$ when $\alpha > 0,\,\delta > 0,\,\theta-\xi > 0$ or when $\alpha > 0,\,\delta < 0,\,\theta - \xi < 0$, \label{en:ineq1}
\item $R_x>R_y$ and $R_z>R_y$ when $\alpha > 0,\,\delta > 0,\,\theta - \xi < 0$ or when $\alpha > 0,\,\delta < 0,\,\theta - \xi > 0$, \label{en:ineq2}
\item $R_y>R_x$ and $R_y>R_z$ when $\alpha < 0,\,\delta > 0,\,\theta - \xi > 0$ or when $\alpha < 0,\,\delta < 0,\,\theta - \xi < 0$, \label{en:ineq3}
\item $R_z>R_y>R_x$ when $\alpha < 0,\,\delta > 0,\,\theta - \xi < 0$ or when $\alpha < 0,\,\delta < 0,\,\theta - \xi > 0$. \label{en:ineq4}
\end{enumerate}
Note that the signs of $\alpha$, $\delta$, and $\theta-\xi$ cannot conclusively determine an inequality or equality relating $R_x$ and $R_z$ in the scenarios~\ref{en:ineq2} and \ref{en:ineq3}.

\begin{figure*}[h]
\centering
\includegraphics[width=\linewidth]{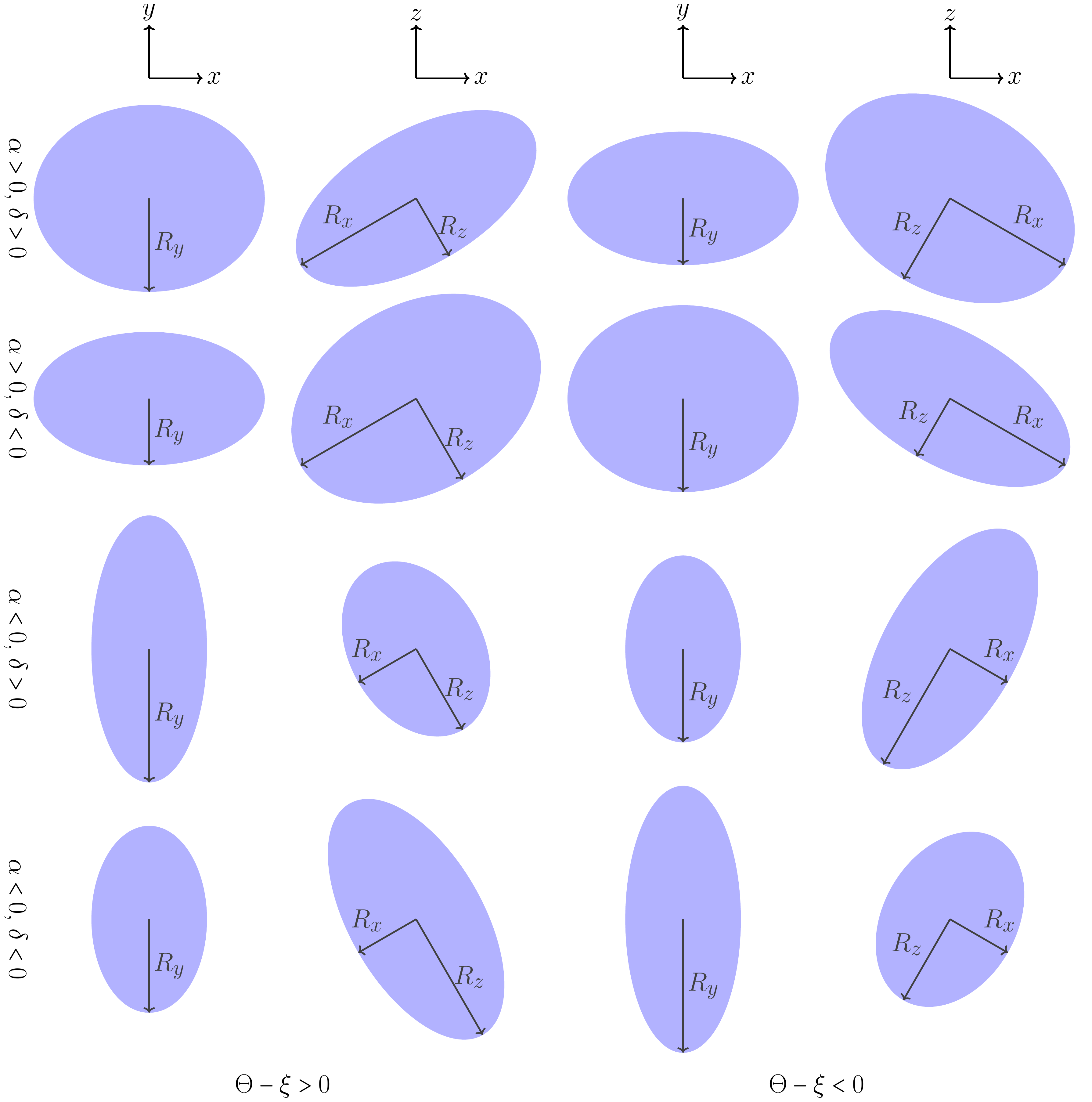}
\caption{Shaded examples of cross-sections at constant $z = 0$ (first and third columns) and $y = 0$ (second and fourth columns), where the TF semi-axes are related to each other in the ratio $5:7:10$. Here, $\theta - \xi$ is positive (first and second columns) or negative (third and fourth columns), $\alpha$ is positive (first and second rows) or negative (third and fourth rows) and $\delta$ is positive (first and third rows) or negative (second and fourth rows).}
\label{crosssecvisual}
\end{figure*}

In Fig.~\ref{crosssecvisual} we illustrate these relations by providing examples of the typical cross-sections of the TF density in the upright co-rotating $x$-$y$ (first and third columns) and $x$-$z$ (second and fourth columns) planes. Each row of Fig.~\ref{crosssecvisual} corresponds to a different combination of positive or negative values of $\alpha$ and $\delta$, with $\alpha > 0,\,\delta > 0$ presented in the first row, $\alpha > 0,\,\delta < 0$ in the second row, $\alpha < 0,\,\delta > 0$ in the third row, and $\alpha < 0,\,\delta < 0$ in the fourth row. In addition, the first and second columns correspond to $\theta - \xi > 0$ and the third and fourth columns correspond to $\theta - \xi < 0$.

\section{\label{sec:level9}Additional GPE Data}
\begin{figure}[H]
\includegraphics[width=\linewidth]{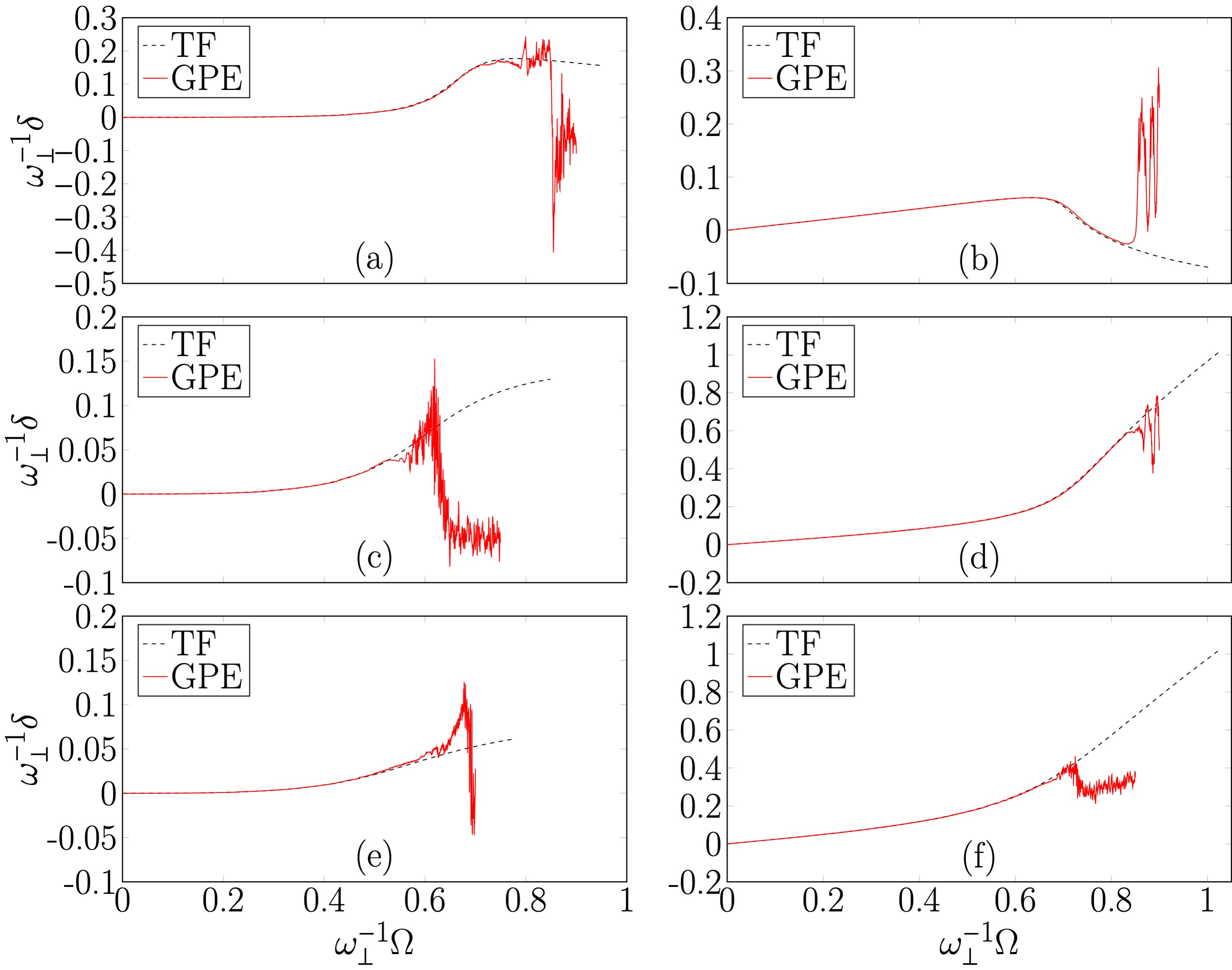}
\vspace*{-5mm}
\caption{Comparison of $\delta$, as a function of $\Omega$, between the TF stationary solution values along Branch I and those derived from GPE simulations of a quasi-adiabatic rampup of $\Omega$, for $\varepsilon = 0$ and $\gamma = 3/4$ (a, c, e) and $\varepsilon = 0.05$ and $\gamma = 4/3$ (b, d, f). $\theta$ equals $\pi/8$ in the first row (a, b), $\pi/4$ in the second row (c, d), or $3\pi/8$ in the third row (e, f).}
\label{cases_1and2_delta}
\end{figure}

For the sake of completeness, we present a comparison of the values of $\delta$, between those pertaining to the TF stationary solutions and those obtained from the GPE simulations in Fig.~\ref{cases_1and2_delta}. In this figure, the first column (a, c, e) pertains to the trapping parameters $\varepsilon = 0$ and $\gamma = 3/4$, and in the second column (b, d, f), $\varepsilon = 0.05$ and $\gamma = 4/3$. The trap tilt angles represented in the Fig.~\ref{cases_1and2_delta} are $\theta = \pi/8$ (first row), $\theta = \pi/4$ (second row), and $\theta = 3\pi/8$ (third row); the analogous comparisons of $\alpha$ and $\varepsilon$ are found in Figs.~\ref{case_1_alphatmx} and \ref{case_2_alphatmx} for the parameters represented in the first and second columns, respectively. The deviation of the GPE-derived values of $\delta$ from the corresponding stationary state values illustrates the transition from the TF state to that with vortices, due to the dynamical instability of the TF states as $\Omega\rightarrow\min\lbrace\Omega_{xc}, \Omega_{yc}\rbrace$, that is discussed in the main text.

\begin{figure}[H]
  \includegraphics[width=\linewidth]{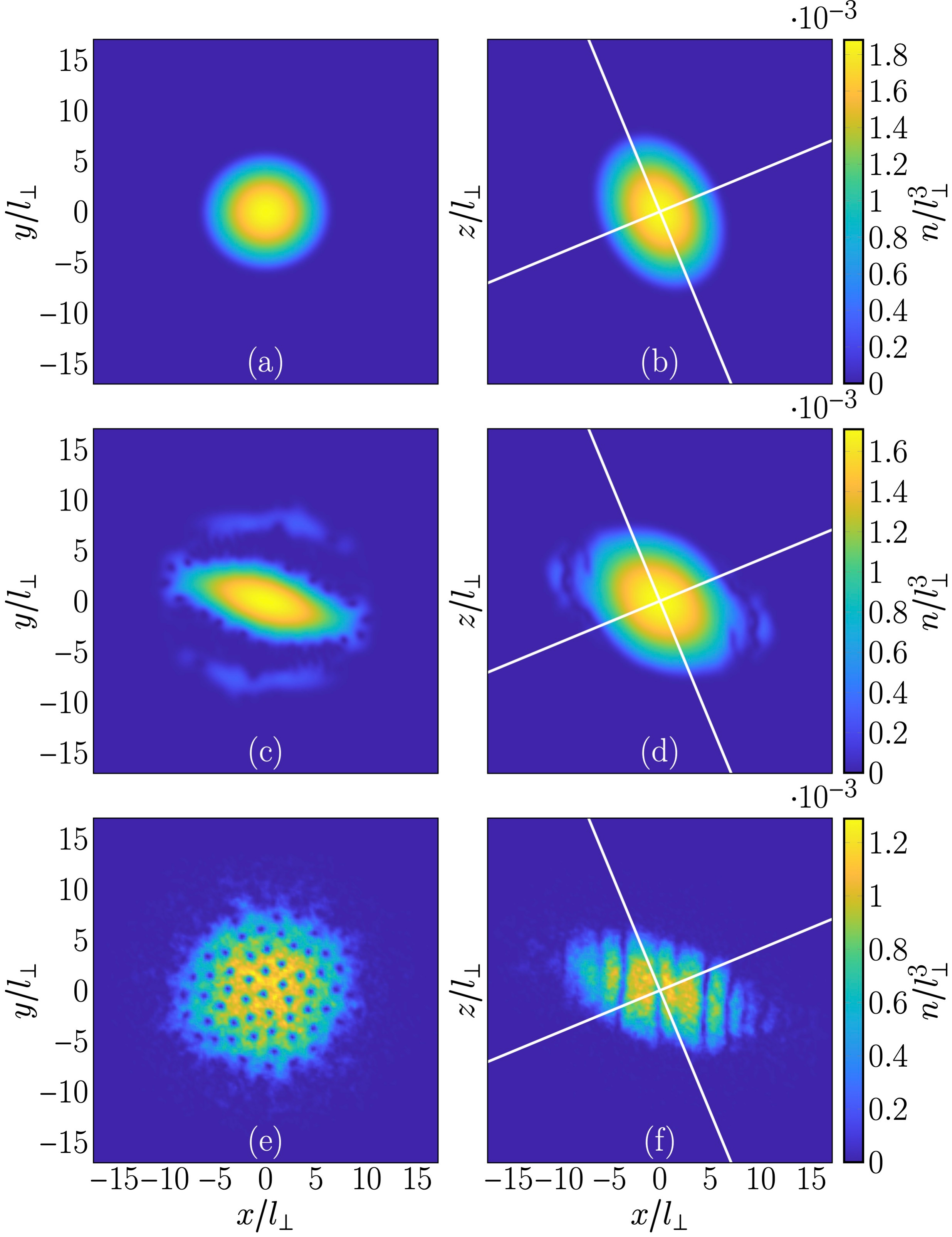}
  \vspace*{-5mm}
  \caption{Cross-sections of the condensate density in the co-rotating $x$-$y$ (first column) and $x$-$z$ (second column) planes for $\varepsilon = 0$, $\gamma = 3/4$, and $\theta = \pi/8$, during a quasi-adiabatic rampup of $\Omega$ at $\Omega = 0.5\omega_{\perp}$ (first row) and $\Omega = 0.8\omega_{\perp}$ (second row), and after $500\omega_{\perp}^{-1}$ at constant $\Omega = 0.8\omega_{\perp}$ (third row). The white lines represent the co-rotating $X$-$Z$ axes.}
  \label{case1_22point5}
\end{figure}

\begin{figure}[H]
  \includegraphics[width=\linewidth]{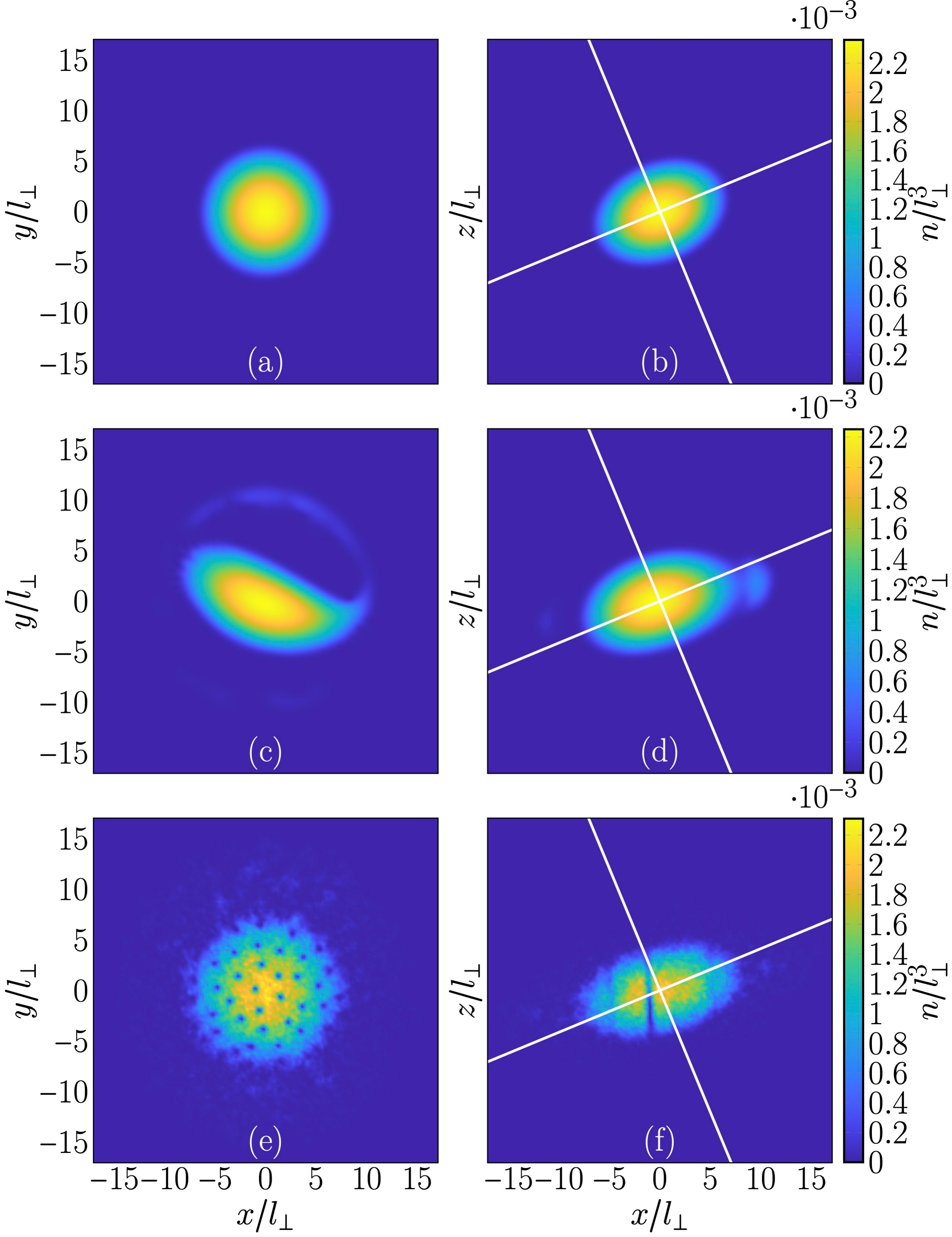}
  \vspace*{-5mm}
  \caption{Cross-sections of the condensate density in the co-rotating $x$-$y$ (first column) and $x$-$z$ (second column) planes for $\varepsilon = 0.05$, $\gamma = 4/3$, and $\theta = \pi/8$, during a quasi-adiabatic rampup of $\Omega$ at $\Omega = 0.5\omega_{\perp}$ (first row) and $\Omega = 0.85\omega_{\perp}$ (second row), and after $500\omega_{\perp}^{-1}$ at constant $\Omega = 0.85\omega_{\perp}$ (third row). The white lines represent the co-rotating $X$-$Z$ axes.}
  \label{case2_22point5}
\end{figure}

\begin{figure}[H]
  \includegraphics[width=\linewidth]{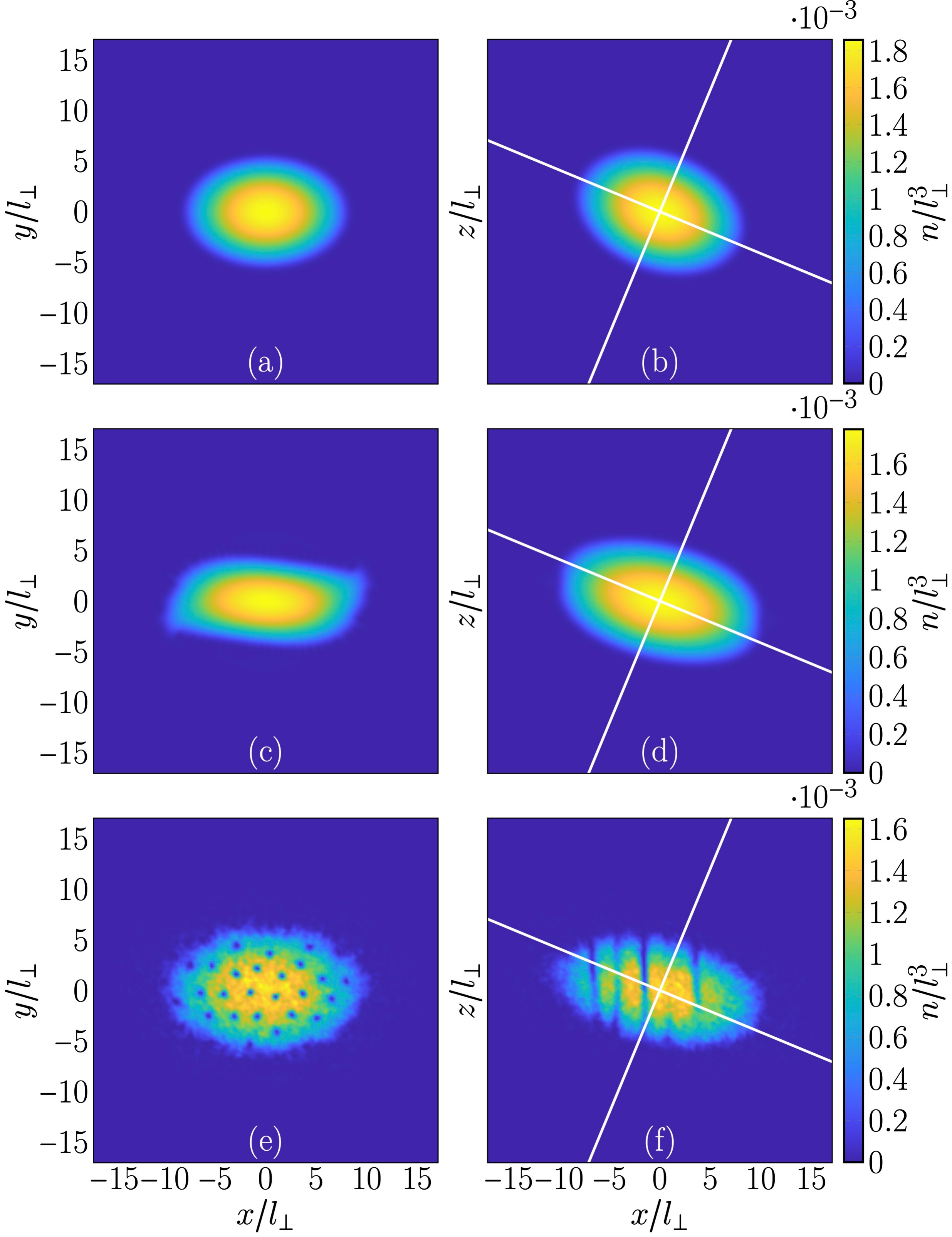}
  \vspace*{-5mm}
  \caption{Cross-sections of the condensate density in the co-rotating $x$-$y$ (first column) and $x$-$z$ (second column) planes for $\varepsilon = 0$, $\gamma = 3/4$, and $\theta = 3\pi/8$, during a quasi-adiabatic rampup of $\Omega$ at $\Omega = 0.4\omega_{\perp}$ (first row) and $\Omega = 0.6\omega_{\perp}$ (second row), and after $500\omega_{\perp}^{-1}$ at constant $\Omega = 0.6\omega_{\perp}$ (third row). The white lines represent the co-rotating $X$-$Z$ axes.}
  \label{case1_67point5}
\end{figure}

\begin{figure}[H]
  \includegraphics[width=\linewidth]{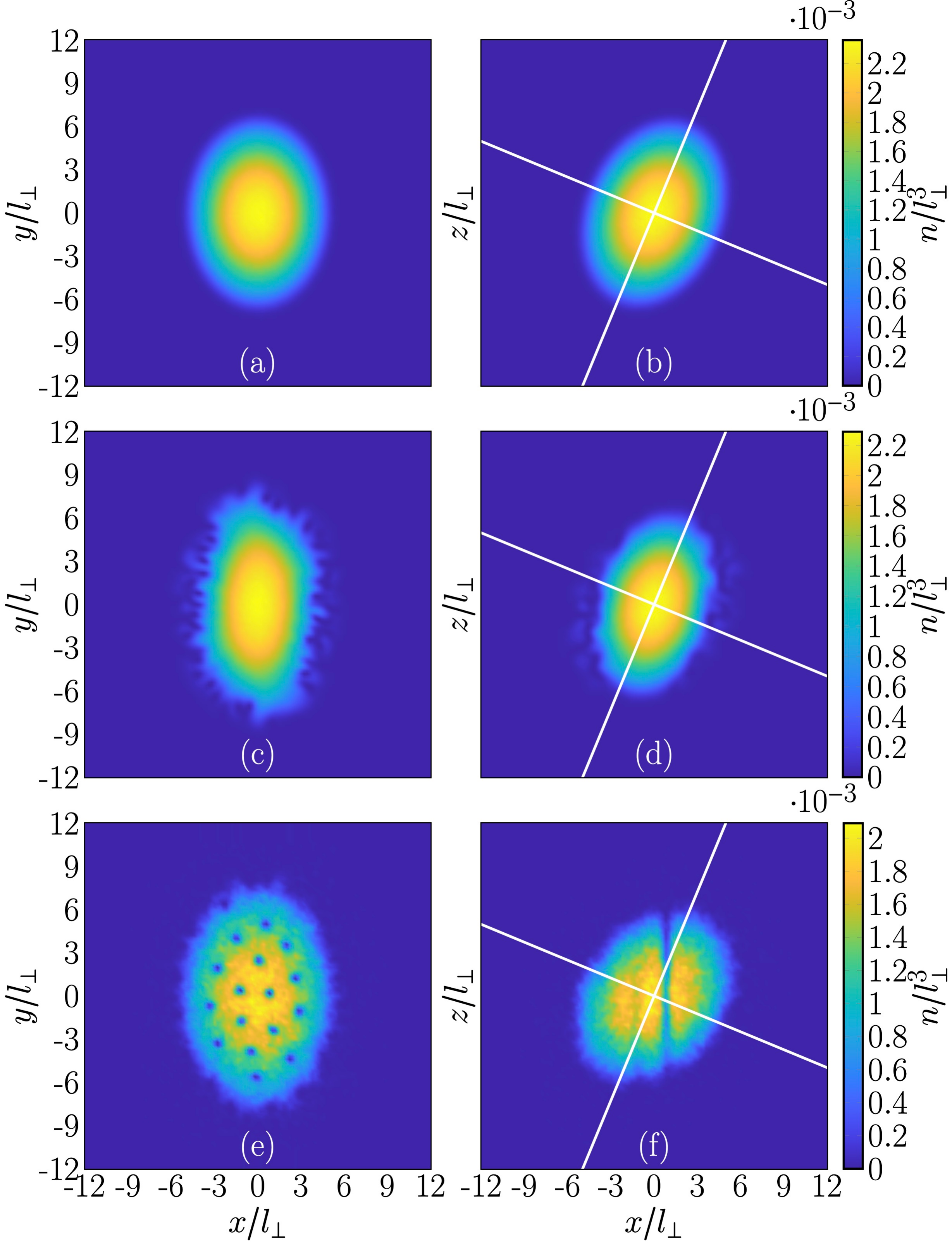}
  \vspace*{-5mm}
  \caption{Cross-sections of the condensate density in the co-rotating $x$-$y$ (first column) and $x$-$z$ (second column) planes for $\varepsilon = 0.05$, $\gamma = 4/3$, and $\theta = 3\pi/8$, during a quasi-adiabatic rampup of $\Omega$ at $\Omega = 0.4\omega_{\perp}$ (first row) and $\Omega = 0.7\omega_{\perp}$ (second row), and after $500\omega_{\perp}^{-1}$ at constant $\Omega = 0.7\omega_{\perp}$ (third row). The white lines represent the co-rotating $X$-$Z$ axes.}
  \label{case2_67point5}
\end{figure}

We also present the cross-sections of the density, in the $x$-$y$ and $x$-$z$ planes, for the angles $\theta = \lbrace \pi/8, 3\pi/8\rbrace$ that were not discussed in the main text. Specifically, for $\theta = \pi/8$, these GPE-derived density cross-sections are plotted in Fig.~\ref{case1_22point5} for the trap with $\varepsilon = 0$ and $\gamma = 3/4$ and in Fig.~\ref{case2_22point5} for the parameters $\varepsilon = 0.05$ and $\gamma = 4/3$. Similarly, for $\theta = 3\pi/8$, the density cross-sections are presented in Fig.~\ref{case1_67point5} for the trapping parameters $\varepsilon = 0$ and $\gamma = 3/4$ and in Fig.~\ref{case2_67point5} for the parameters $\varepsilon = 0.05$ and $\gamma = 4/3$. Note that the values of $\Omega$ where the GPE cross-section snapshots are taken have chosen in order to illustrate the three main stages of the evolution of the BEC from TF-like, via the intermediate stage with a halo-like cloud surrounding the deformed core, to a state containing many vortices.

\bibliography{main.bbl}
\end{document}